\documentclass[11pt,letterpaper]{article}
\pdfoutput=1
\usepackage{etoolbox}
\makeatletter
\patchcmd{\maketitle}{\@fpheader}{}{}{}
\makeatother

\usepackage{jheppub}
\usepackage{color}
\usepackage{wrapfig,enumerate,slashed}
\usepackage{graphicx,epsf,bm}
\usepackage{amssymb}
\usepackage{amsmath}
\usepackage{epstopdf}
\usepackage{booktabs}
\usepackage{subfigure}
\usepackage{rotating}
\usepackage{xspace}

\usepackage{footmisc}
\usepackage{amsmath}
\usepackage{wasysym} 
\usepackage{graphicx}
\usepackage{color}
\usepackage{comment}
\usepackage{hyperref}

\newcommand{\heptt}{HEPTopTagger\xspace}

\newcommand{\hptt}{HPTTopTagger\xspace}

\newcommand{\pythia}{PYTHIA\xspace}
\newcommand{\sherpa}{SHERPA\xspace}
\newcommand{\herwigpp}{HERWIG++\xspace}

{}
{}
{}

\newcommand{\delphes}{Delphes\xspace}
\newcommand{\fastjet}{FastJet\xspace}
\newcommand{\figref}[1]{Figure~\ref{fig:#1}}
\newcommand{\secref}[1]{Section~\ref{sec:#1}}
\renewcommand{\eqref}[1]{Equation~(\ref{eq:#1})}
\newcommand{\tabref}[1]{Table~\ref{tab:#1}}

\newcommand{\MeV}{\ensuremath{\,\text{MeV}}}
\newcommand{\GeV}{\ensuremath{\,\text{GeV}}}
\newcommand{\TeV}{\ensuremath{\,\text{TeV}}}
\newcommand{\invfb}{\ensuremath{\,\text{fb}^{-1}}}

\begin{document}

\title{\bf \Large Tracking New Physics at the LHC and beyond}

\author[a]{Michael Spannowsky}
\author[b]{and Martin Stoll}

\affiliation[a]{Institute for Particle Physics Phenomenology,
  University of Durham, Durham DH1 3LE, UK}
\affiliation[b]{Department of Physics, University of Tokyo, Bunkyo-ku, Tokyo 113-0033, Japan}

\emailAdd{michael.spannowsky@durham.ac.uk}
\emailAdd{stoll@hep-th.phys.s.u-tokyo.ac.jp}

\abstract{
Heavy resonances are an integral part of many extensions of the Standard Model. The discovery of such heavy resonances are a primary goal at the LHC and future hadron colliders.
When a particle with TeV-scale mass decays into electroweak-scale objects, these objects are highly boosted and their decay products are then strongly collimated, possibly to an extent that they cannot be resolved in the calorimeters of the detectors any more. We develop taggers for electroweak-scale resonances by combining the good energy resolution of the hadronic calorimeter with the superior spatial resolution of the tracking detector. Using track-based techniques we reconstruct heavy $W'$ and $Z'$ bosons and constrain the branching ratio of the rare Higgs boson decay $H \to Z A \to l^+l^-$ jets.
The taggers show a good momentum-independent performance up to very large boosts. Using the proposed techniques will allow experiments at the LHC and a future hadron collider to significantly extend its reach in searches for heavy resonances.
}

\preprint{IPPP/14/12,DCPT/14/24,UT-15-11}

\def\thepage{{}}
\maketitle
\def\thepage{\arabic{page}}

\section{Introduction}
\label{sec:case}
After the successful discovery of the Higgs boson during LHC run 1 \cite{Aad:2012tfa,Chatrchyan:2012ufa}, the next foremost goal of the upcoming runs is to discover new physics, e.g.~new heavy resonances. The scale of new physics for many anticipated extensions of the Standard Model has already been pushed beyond $\mathcal{O}(1)$ TeV \cite{Aad:2015pfa,Aad:2014xea, Khachatryan:2015sja,Khachatryan:2015lwa}. If heavy TeV-scale resonances decay into electroweak-scale particles which in turn have large branching ratios into quarks, i.e.~$X_{\mathrm{TeV}} \to Y_{\mathrm{EW}} \to \mathrm{jets}$, these quarks are likely to be collimated in the lab frame. 
More precisely, the extent to which the decay products of the electroweak-scale resonances are collimated depends on the ratio between the mass of the heavy new-physics resonance and the electroweak scale. For central production one finds $p_{\perp,Y} \sim m_X/2 $. As a result, for the decay products of Y their angular separation scales like $\Delta R_\mathrm{jets} \sim 4 m_Y/m_X$.
As electroweak-scale resonances have generically large branching ratios into quarks the reconstruction and detailed analysis of hadronic final states is at the core of the upcoming LHC program.

In an experiment jets are reconstructed using infrared-safe jet algorithms \cite{Ellis:1993tq,Catani:1993hr,ca_algo,Cacciari:2008gp}. Input to jet algorithms are so-called topo-clusters, objects constructed from long-lived particles' energy deposits in the electromagnetic (e-cal) and hadronic (h-cal) calorimeter \cite{LopezMateos:2011fta}. 
The minimal transverse size for a cluster of hadronic calorimeter cells is $0.3 \times 0.3$ in $(\eta, \phi)$, reached if all energy after noise-subtraction is concentrated in one cell. To discriminate two jets the angular separation of their axes in the detector has to be at least $\Delta R \equiv \sqrt{(\Delta \eta)^2 +  (\Delta \phi)^2} = 0.2$. For most Standard Model processes at the LHC this angular resolution is sufficient to separate the decay products of electroweak resonances. However, when scales are vastly separated and either $4 m_Y/m_X \ll 0.2$ or in general $p_{\perp,X} \gg m_X$ the angular separation of the decay products can be too small to resolve them individually. While the overall energy deposit of highly boosted resonances can still be measured, the substructure, i.e.~the energy sharing between the decay products, becomes oblivious and the ability to discriminate between a decaying resonance and QCD jets using jet substructure observables quickly deteriorates. Obviously, at a possible future 100 TeV proton-proton collider where larger $m_X$ are probed and the rate for electroweak resonances with $p_{\perp,X} \gg m_X$ is bigger, this issue cannot be ignored.

To recover reconstruction efficiency for highly boosted resonances and extend the multi-purpose experiments' sensitivity in searches for heavy resonances to larger masses, smaller input objects to jet algorithms have to be used. The most popular taggers using jet substructure are either based on jet-shape observables~\cite{thaler_wang, leandro1, nsub,treeless} or on calibrated subjets  inside a fat jet~\cite{hopkins,pruning1, cmstagger,heptop, Soper:2012pb}. 

 Following \cite{hpttop}, we propose to use tracks instead of topo-clusters as input for resonance reconstruction methods\footnote{In an earlier proposal \cite{Katz:2010mr} the same approach for the electromagnetic calorimetry was discussed for $W$ tagging.} and we develop and refine dedicated reconstruction procedures for highly-boosted electroweak-scale resonances, e.g.~$W$/$Z$ bosons and top quarks. While we will focus on subjet based techniques the same approach can be used for jet shape observables \cite{Larkoski:2015yqa}.

The article is arranged as follows:
In \secref{toptag} we briefly describe the top-tagging algorithm dubbed High-pT TopTagger (\hptt) and discuss modifications regarding the proposal in \cite{hpttop}. 
In \secref{wztag} we extend the concept of the \hptt to $W$ and $Z$ boson tagging and apply these taggers in heavy resonance searches at the LHC and a 100 TeV FCC-hh. 
In \secref{higgstag} we focus on a rare Higgs decay and show that in searches for very light resonances track-based reconstruction can be an indispensable tool in the upcoming LHC runs. Eventually we summarize our findings in \secref{outlook}.

\section{Tagging highly boosted top quarks}
\label{sec:toptag}

\subsection{The new default of the \hptt}
\label{sec:toptag:algorithm}

The \hptt of Ref.~\cite{hpttop} reconstructs the three-prong substructure of hadronically decaying top quarks from charged tracks. These particle trajectories can be determined in a tracking detector to very high radial precision.
More specifically, the ATLAS inner tracking detector achieves an angular reslution of $\Delta \eta \approx 10^{-3}$ and $\Delta \phi \approx 0.3\,\text{mrad}$ for charged particles with $p_\perp=10\GeV$ \cite{Aad:2008zzm}, while maintaining a reconstruction efficiency of $>78\%$ for tracks of charged particles with $p_\perp\geq 500\MeV$ \cite{Aad:2010ac}. 
Such accuracy cannot be achieved with the hadronic calorimeter where it is not possible to separately resolve small jets with $\Delta R_{j_1,j_2} < 0.2$. 

Tagging algorithms that rely solely on information from calorimeter towers on the other hand -- such as e.g.~the \heptt \ \cite{heptop} -- are therefore not applicable any more if the top daughter jets are strongly collimated.
The track-based \hptt ports elements of the HEPTopTagger algorithm to the high-energy regime. However, to avoid combinatorial issues due to a high multiplicity of tracks and the introduction of artificial mass scales in background events, cuts are only applied on one three-subjet configuration in a large-radius fat jet. Hence we do not search for a top-like structure in every possible subjet combination.
The bulk of the top identification is then achieved by comparing ratios of invariant mass combinations of the three subjets. For example, the ratio $m_{23}/m_{123}$ corresponds to $m_W/m_t$ in most hadronic top decays, where $m_{23}$ is the invariant mass of the sub-leading and sub-sub-leading subjet in transverse momentum and $m_{123}$ is the invariant mass of the top candidate.

Due to imperfect knowledge of all energy flow in the tracking detector (only charged particles are reconstructed), the track momenta are scaled according to the inverse of the energy fraction carried by charged tracks \cite{Katz:2010mr,hpttop}
\begin{align}
\label{eq:alpha}
 \alpha_j \equiv \frac{E_\text{jet}}{E_\text{tracks}} \,.
\end{align}
Because the energy of the (hadronic) fat jet can be calibrated to good precision in the experiment \cite{Aad:2012vm,Aad:2013gja}, the sensitivity to fluctuations is hence ameliorated. In essence we propose to perform a local recalibration of track-based (sub)jets, according to the relative energy deposit of tracker and calorimetry. Thereby we seek to benefit from the tracker's spatial and the calorimeter's improved energy resolution.

\begin{figure}[ht]
\begin{center}
 \includegraphics[width=0.7\textwidth, trim=5cm 14cm 3cm 4cm, clip=true]{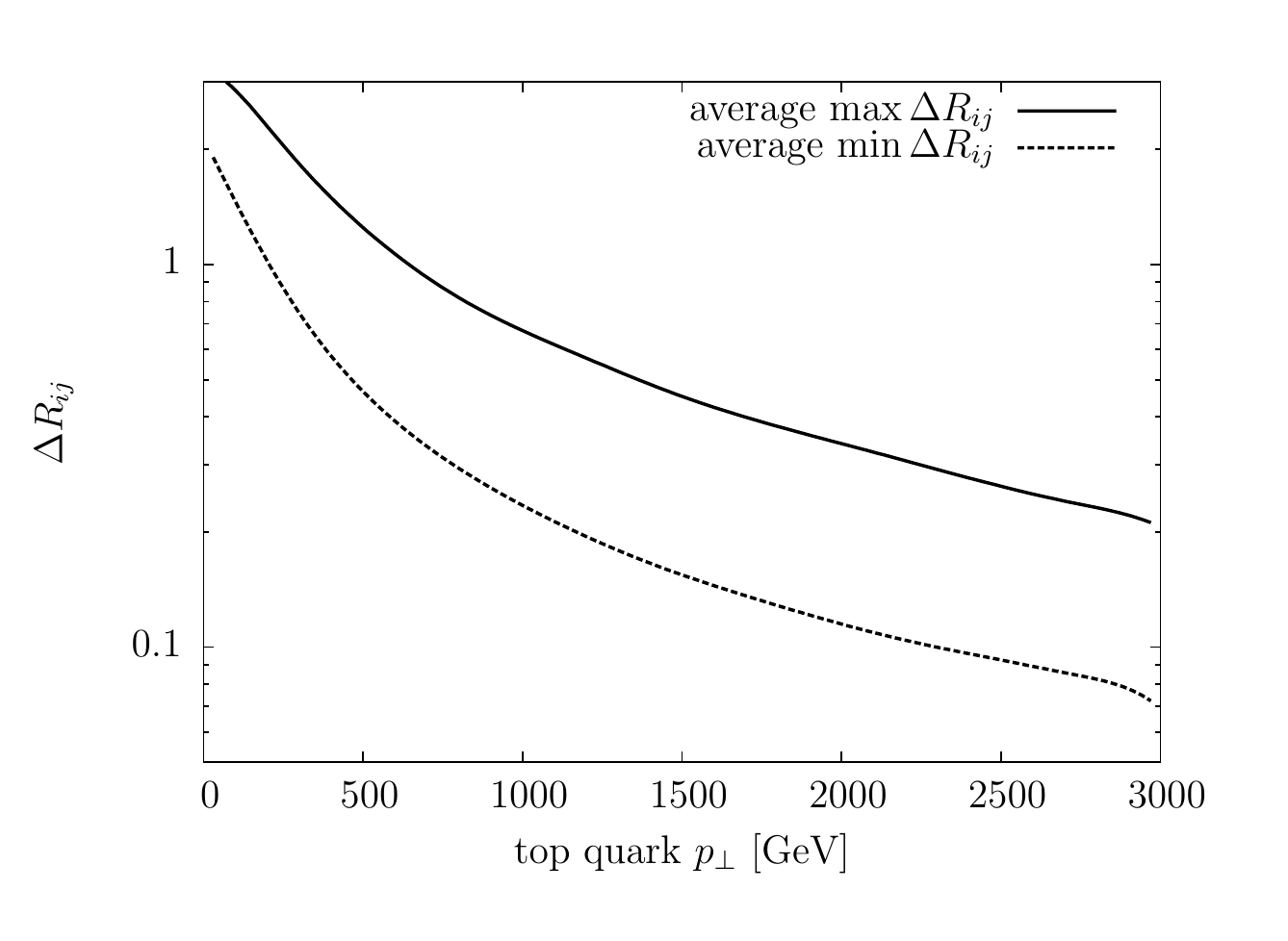}
 \caption{Average smallest and largest mutual separation in $\Delta R_{ij} \equiv \sqrt{\Delta\phi_{ij}^2+\Delta\eta_{ij}^2}$ between the partons from the hadronic top decay $t\to Wb\to jjb$.}
 \label{fig:top_size}
\end{center}
\end{figure}

In fact, calibrations of Cambridge-Aachen (C/A) \cite{ca_algo} subjets are available for radius parameters as small as 0.2 \cite{Aad:2013gja} which makes it beneficial to re-scale the charged tracks more locally.
For boosted tops with transverse momentum at the TeV scale, typically two or all three top subjets can be resolved in this way and the fluctuations are expected to be reduced separately.
In \figref{top_size} we show the average mutual separation between top decay products.
Thus, among other improvements, the new new default of the HPTTopTagger applies the approach of \eqref{alpha} to subjets individually instead of the whole fat jet. 

\begin{figure}[ht]
\begin{center}
 \includegraphics[width=0.45\textwidth]{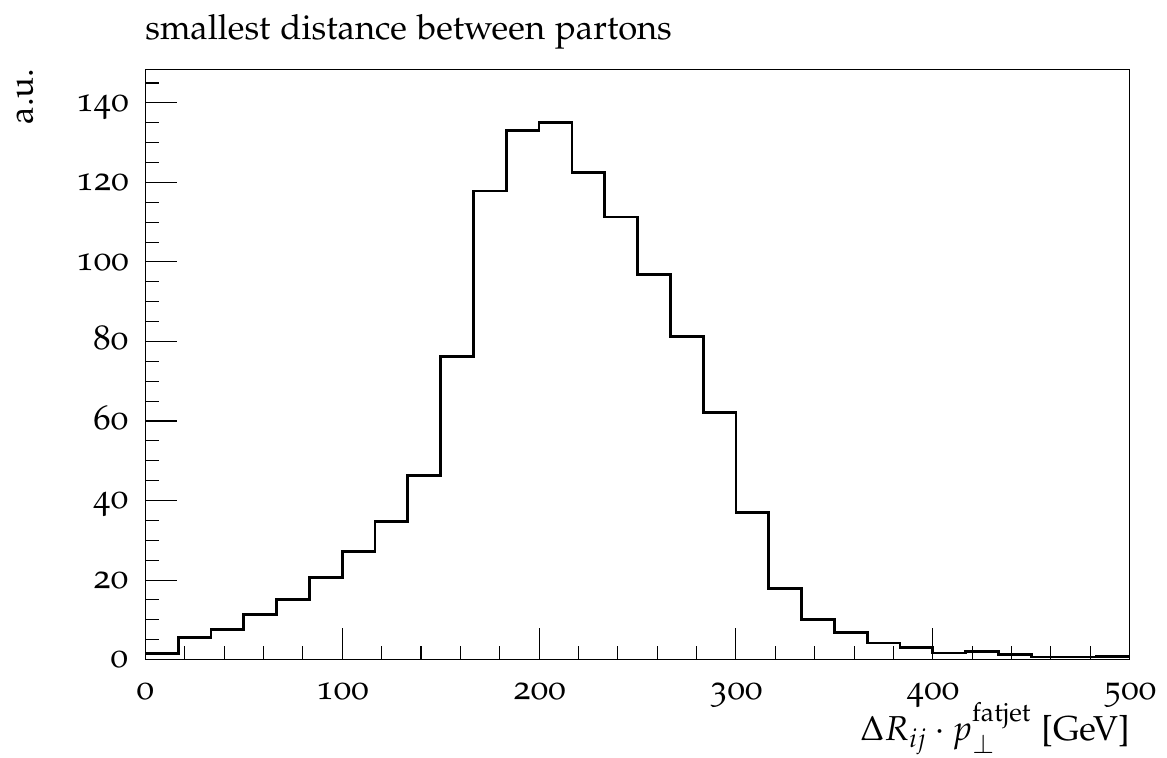}
 \includegraphics[width=0.45\textwidth]{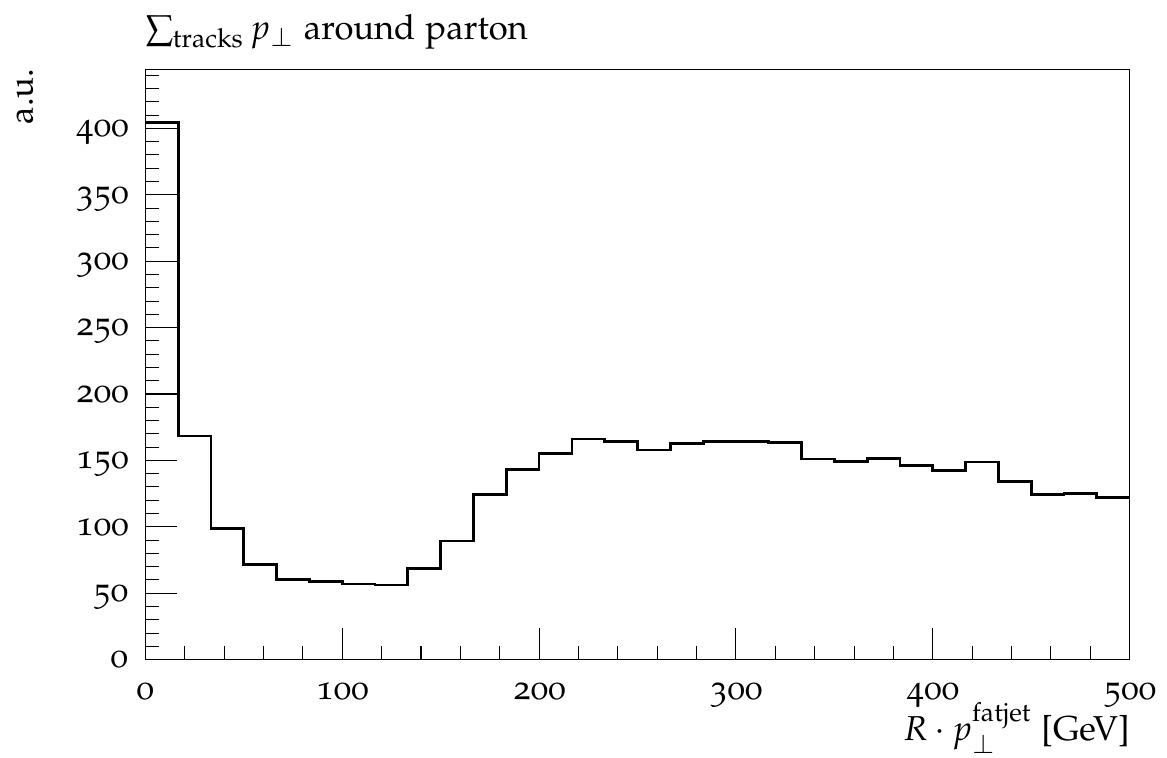}
 \caption{Smallest separation between two MC truth partons in a top decay, scaled with the fat jet transverse momentum (left) and distribution of transverse momentum carried by charged tracks around the truth partons (right).}
 \label{fig:sub_separation}
 \end{center}
\end{figure}

We find that the Monte Carlo truth partons from the decay $t\to Wb\to jjb$ are separated by a characteristic $R$ distance of $\sim 200\GeV/p_{\perp,t}$ and that  the energy carried by charged tracks around these partons is very well localized with a much smaller radius, cf.~\figref{sub_separation}.
Given these observations, we are led to abandon the mass-drop unclustering procedure which was inherited from the \heptt in favour of conventional (anti-$k_T$) \cite{Cacciari:2008gp} subjets with radius parameter $R=100\GeV/p_{\perp,j_c}$.
We label the three subjets leading in transverse momentum $\tilde{j}_1$, $\tilde{j}_2$ and $\tilde{j}_3$.
This procedure renders an additional filtering stage redundant.

The inclusion of the leading gluon emissions is important to capture all top quark decay products, in particular when $p_{\perp,t} \gg m_t$.
In~\cite{hpttop} this was not done explicitly, which resulted in relatively low reconstructed masses. 
Additional soft emissions can however easily be captured if we allow our subjets to overlap, i.e.~we recluster the fat jet $j_c$ with $R=0.8\cdot \min \Delta R(\tilde{j}_i,\tilde{j}_k)$ where $i,k \in  \{1,2,3~|~i\neq k\}$.
As the new radius is smaller than the subjet separation, it is guaranteed that they are still resolved separately.
The new $p_\perp$-leading subjets $j_1$, $j_2$ and $j_3$ form the top candidate.

Typically tagging efficiencies will be larger for the new \hptt which is a desired feature, especially when the luminosity of the experiment is limited. To achieve good rejection rates of background events, in addition to the so-called A-cut of mass ratios, we apply a cut based on the geometric size of the top candidate. The scaled spread of the reconstructed subjets has to satisfy 
\begin{eqnarray}
 \max\Delta R(j_i,j_k) \cdot p_{\perp,\text{top candidate}} \in [\, 250\GeV , 750\GeV\,]
\end{eqnarray}
where $i,k \in  \{1,2,3~|~i\neq k\}$, in order to reject fake candidates where the big top-like mass is generated through soft large-angle radiation.
\newline

We define the new default of the \hptt algorithm by the following procedure:
\begin{enumerate}
 \item Define a fat jet $j$ using the C/A algorithm with $R=0.8$ and $p_\perp\geq800\GeV$.
 \item Discard all tracks that are not associated with $j$ or that have $p_\perp <500\MeV$.
 \item Scale the remaining track momenta as follows: Re-cluster $j$ with the anti-$k_T$ algorithm employing a small radius $R=0.2$, and calculate $\alpha_j\equiv E_\text{jet}/E_\text{tracks}$ for each subjet using its respective associated tracks. We multiply the momenta of the tracks by $\alpha_j$ and recombine the scaled tracks to a track-based jet $j_c$.
 \item Re-cluster $j_c$ using the anti-$k_T$ algorithm with $R=100\GeV/p_{\perp,j_c}$.
 If there are fewer than three subjets we consider the tag to have failed.
 \item We then calculate the smallest pair-wise distance between the three leading subjets,
 $r_\text{min}\equiv\min \Delta R_{ij}$ and re-cluster $j_c$ with a new radius $R=0.8\, r_\text{min}$.
 If the new three leading subjets result in an invariant mass around the top quark mass, $m_\text{candidate}\in m_t\pm 25\GeV$, they form our top candidate.
 \item We follow the \heptt and apply the so-called A-cut of~\cite{heptop} on the pair-wise invariant masses $(m_{12},m_{13},m_{23})$. If in addition the top candidate satisfies
 $p_\perp\cdot \max\Delta R_{ij} \in[\, 250\GeV, 750\GeV \,]$,
 we consider the top tag to be successful.
\end{enumerate}

\subsection{Performance}
\label{sec:toptag:performance}

We investigate the tagging efficiency and the reconstructed mass of the HPTTopTagger, as described in \secref{toptag:algorithm}.
Throughout this section, we generate top-initiated jets from the production of a $Z'$,
$p p\to Z'\to t\bar{t}\to \text{jets}$,
for masses in the range $m_{Z'}=3 \cdots 6\TeV$. 
Unless otherwise stated, the sample with mass $m_{Z'}=6\TeV$ is used to generate the plots.
For the background we consider QCD jets from dijet production.
Unless stated otherwise, we generate events with \pythia~8~\cite{Sjostrand:2007gs} at centre-of-mass energy $\sqrt{s}=14\TeV$.
Fat jets are clustered using \fastjet \cite{fastjet} from stable particles with pseudorapidity $|\eta|<4.9$. We apply the C/A jet algorithm with resolution parameter $R=0.8$ and $p_\perp\geq 800$ GeV. To assess the signal efficiency we match the fat jet to MC truth top quarks by requiring $\Delta R(j,\mathrm{top}) < 0.6$ and select jets with $|\eta_j| < 2.5$.

A priory, constructing observables from tracks can result in a marked sensitivity on detector resolution and efficiencies. To evaluate the impact of these effects on the performance of the algorithm we use the \delphes~\cite{Ovyn:2009tx} fast dectector simulation.
Throughout parameters are chosen in accordance with the default ATLAS parameters provided by \delphes.

\begin{figure}[htb]
 \begin{center}
  \includegraphics[width=0.7\textwidth, trim=5cm 14.5cm 4cm 4.5cm, clip=true]{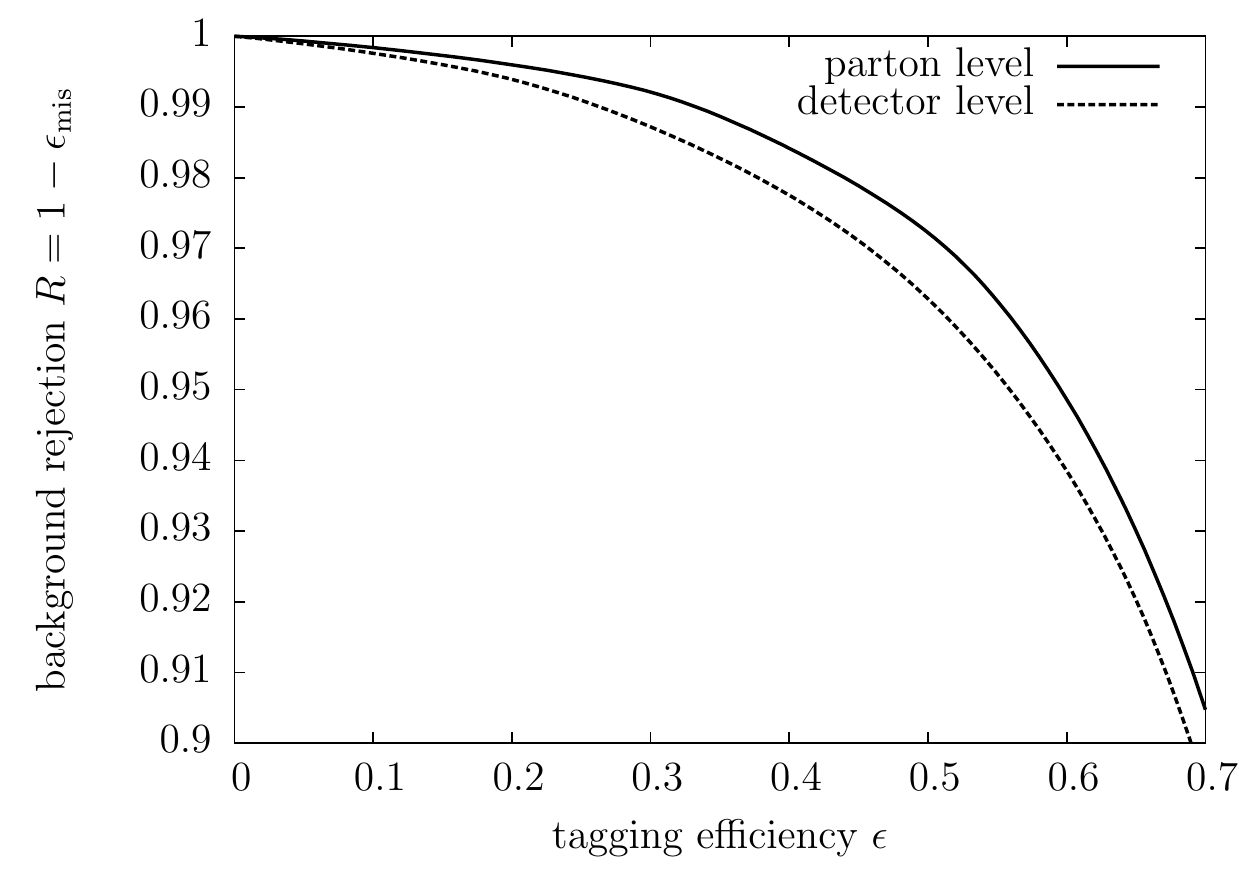}
 \end{center}
 \caption{ROC for scan over $m_{23}/m_{123}$, $p_\perp\cdot \max\Delta R_{ij}$, $m_t$, $m_W$, $\Delta m_t$, $\Delta m_W$ in various combinations and pre-scan cuts. For details on signal and background generation see \secref{toptag:performance}.}
 \label{fig:t_roc}
\end{figure}

\figref{t_roc} shows the receiver-operater-characteristic (ROC) curve for our algorithm.
By tuning the free parameters and cuts, background rejection
$R=1-\epsilon_\text{mis}$
can be maximized for any working point with given signal tagging efficiency $\epsilon$.
We use TMVA \cite{Hocker:2007ht} to find optimum cuts on
$m_{23}/m_{123}$, $p_\perp\cdot \max\Delta R_{ij}$,
as well as the allowed mass windows around $m_t$ and $m_W$.
TMVA assesses the cut parameters by their respective discrimination power and applies cuts in this order.
To further improve performance, we run the algorithm for different combinations of fixed and free cut parameters and combine the resulting curves such that at each working point, the setup with the largest background rejection is selected.
To obtain a dropping $p_\perp$ distribution as expected in a real analysis, background QCD jets we obtain from dijet production with $\hat{p}_\perp\geq 700\GeV$.
$\hat{p}_\perp$ is the transverse momentum in the rest frame of the hard process at generator level.

\begin{figure}[htb]
 \begin{center}
  \includegraphics[width=0.45\textwidth]{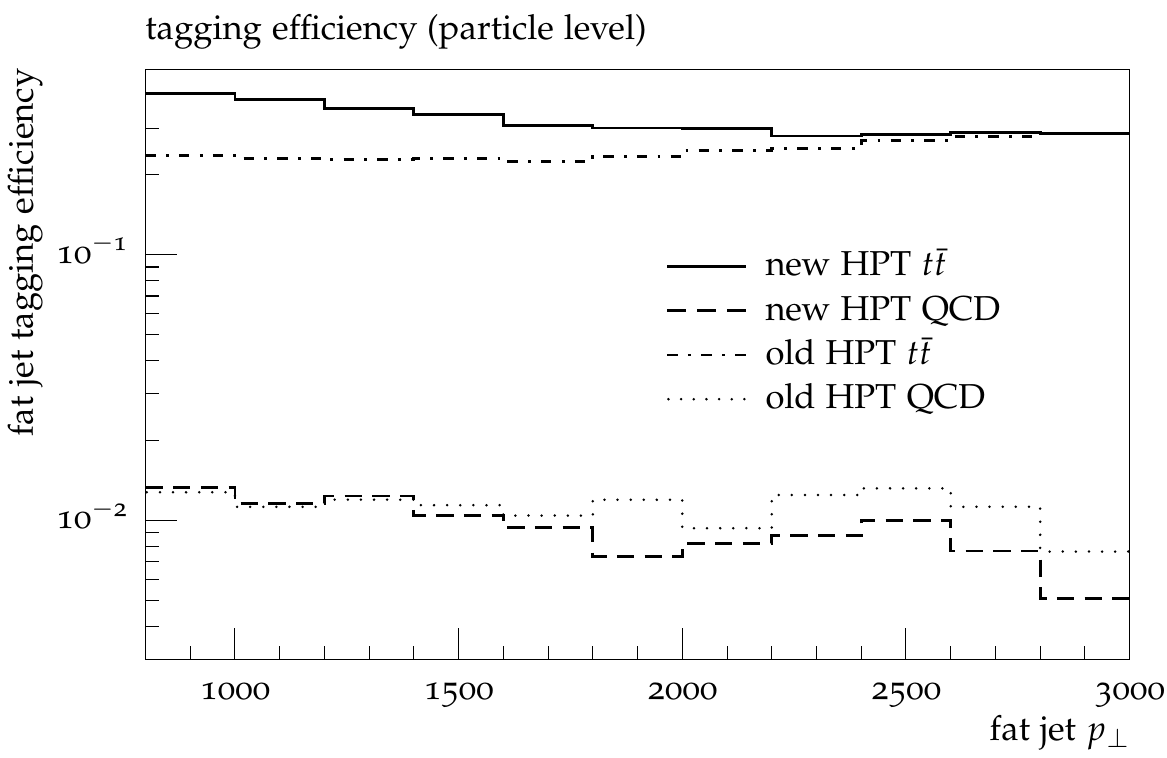}
  \includegraphics[width=0.45\textwidth]{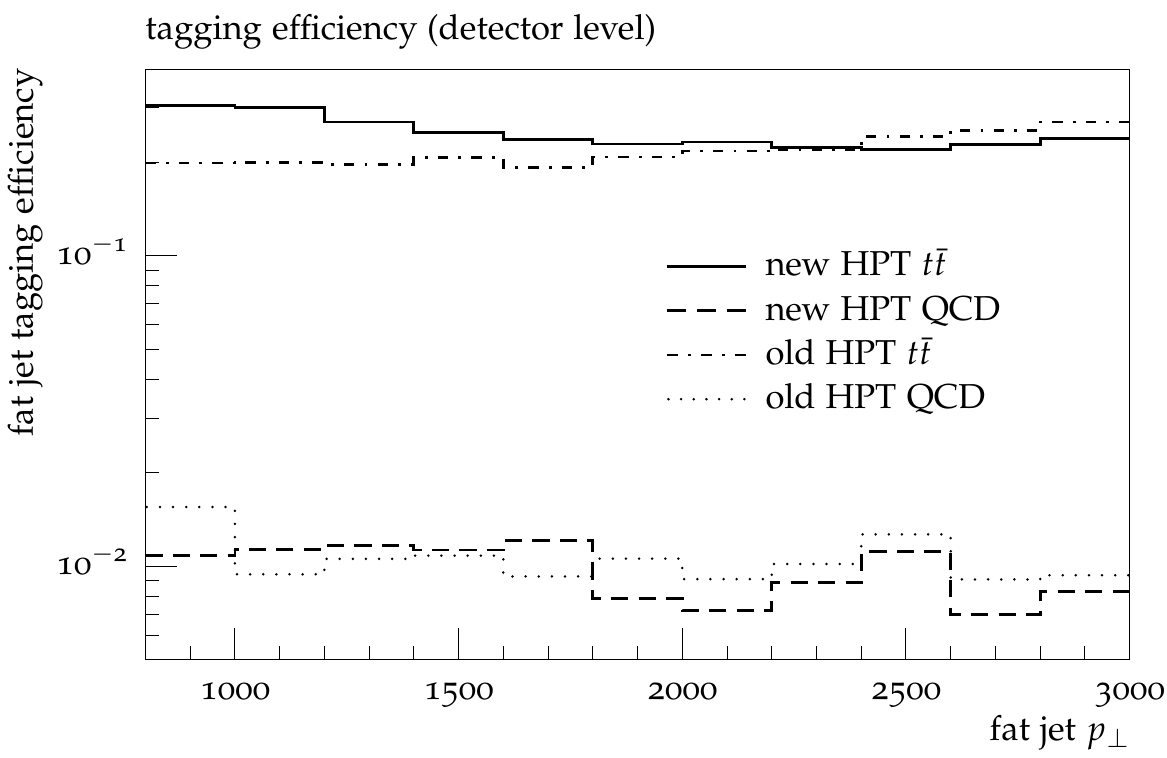}
 \end{center}
 \caption{Tagging efficiency of the \hptt and the new default at particle level (l.h.s.) and detector level (\delphes) (r.h.s.). 
 }
 \label{fig:t_efficiency}
\end{figure}

In \figref{t_efficiency} we show the $p_\perp$-dependent tagging efficiencies of the HPTTopTagger.
Here we obtain background QCD jets from dijet production with binned generator-level $\hat{p}_\perp$ in the range $[700,2500]\GeV$ and $\hat{p}_\perp\geq 2500\GeV$ to achieve good statistics in all bins. 
Over the whole studied $p_\perp$ range we find a flat tagging efficiency and fake rate. The outlined modifications for the new default of the HPTTopTagger improve the tagging efficiency for fat jets with $p_{\perp,t} \sim 1-2$ TeV while maintaining a similar fake rate. For particle-level final states with $p_{\perp,\mathrm{fatjet}} \sim 1$ TeV the HPTTopTagger has a signal efficiency of roughly $35\%$ which slightly decreases to $30\%$ at $p_{\perp,\mathrm{fatjet}} = 3$ TeV. Including detector effects results in a flat shift  to lower values by $5\%$ over the entire $p_{\perp,\mathrm{fatjet}}$ range.

\begin{figure}[htb]
 \begin{center}
  \includegraphics[width=0.45\textwidth]{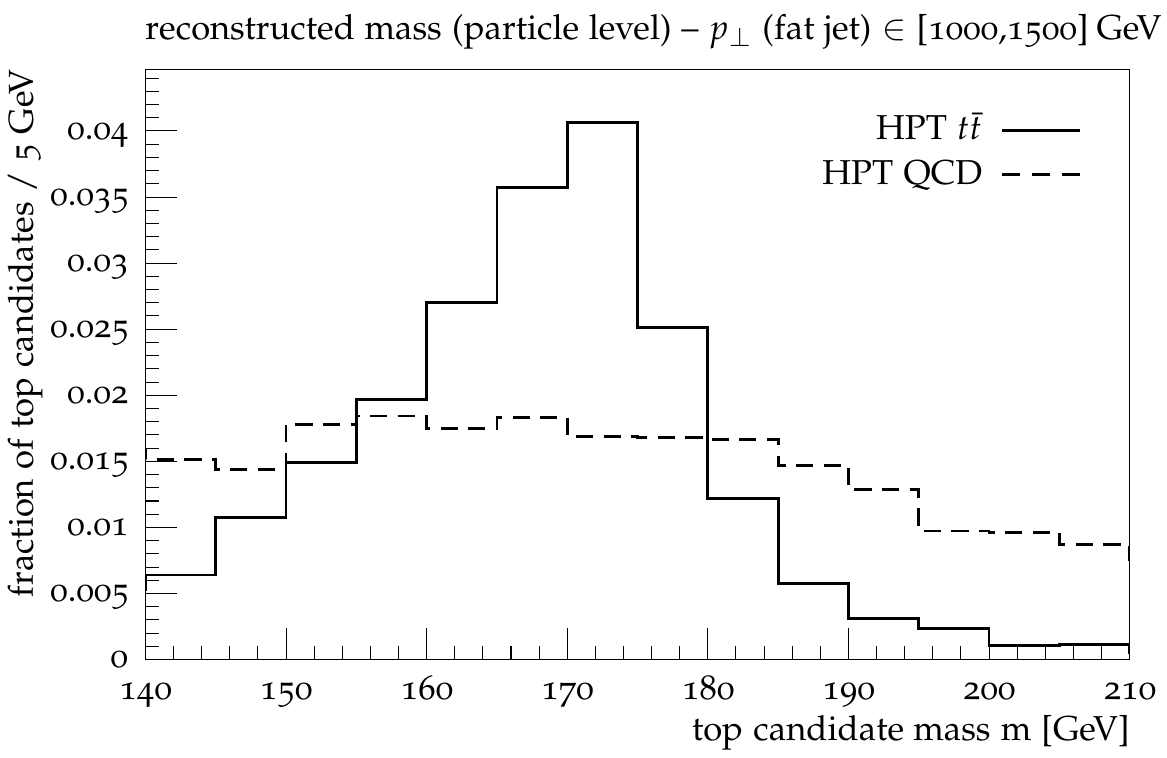}
  \includegraphics[width=0.45\textwidth]{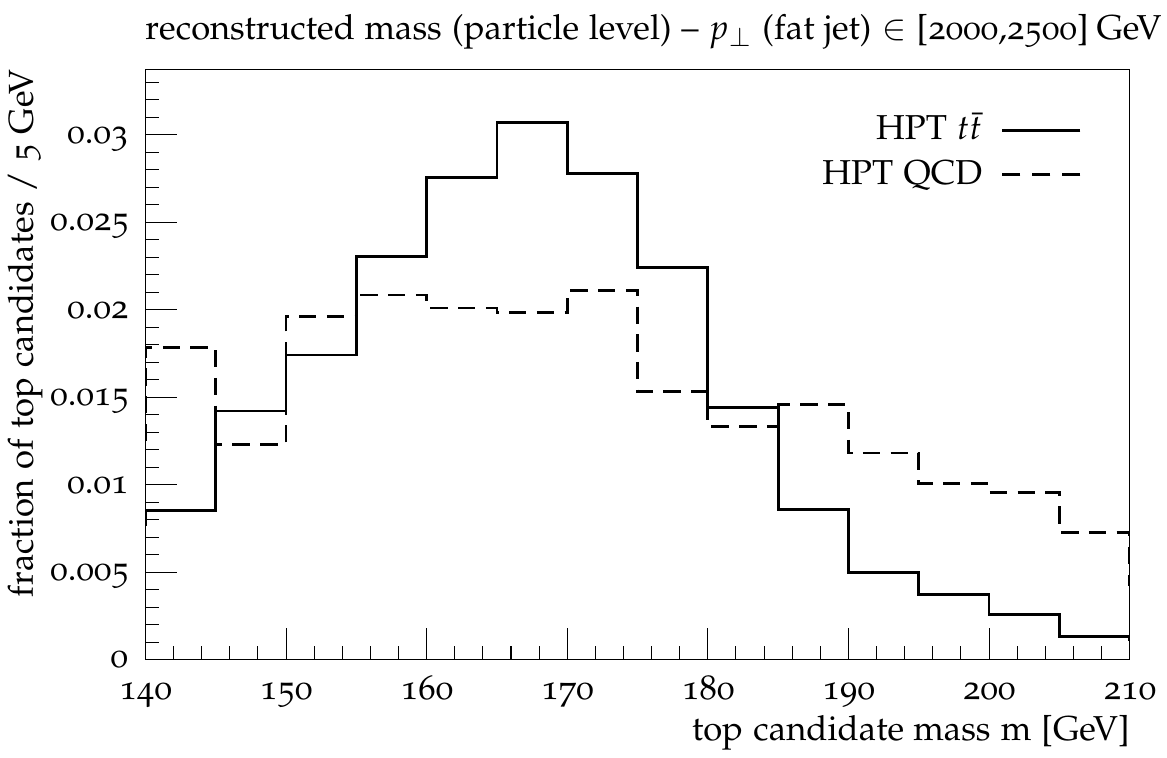}\\
  \includegraphics[width=0.45\textwidth]{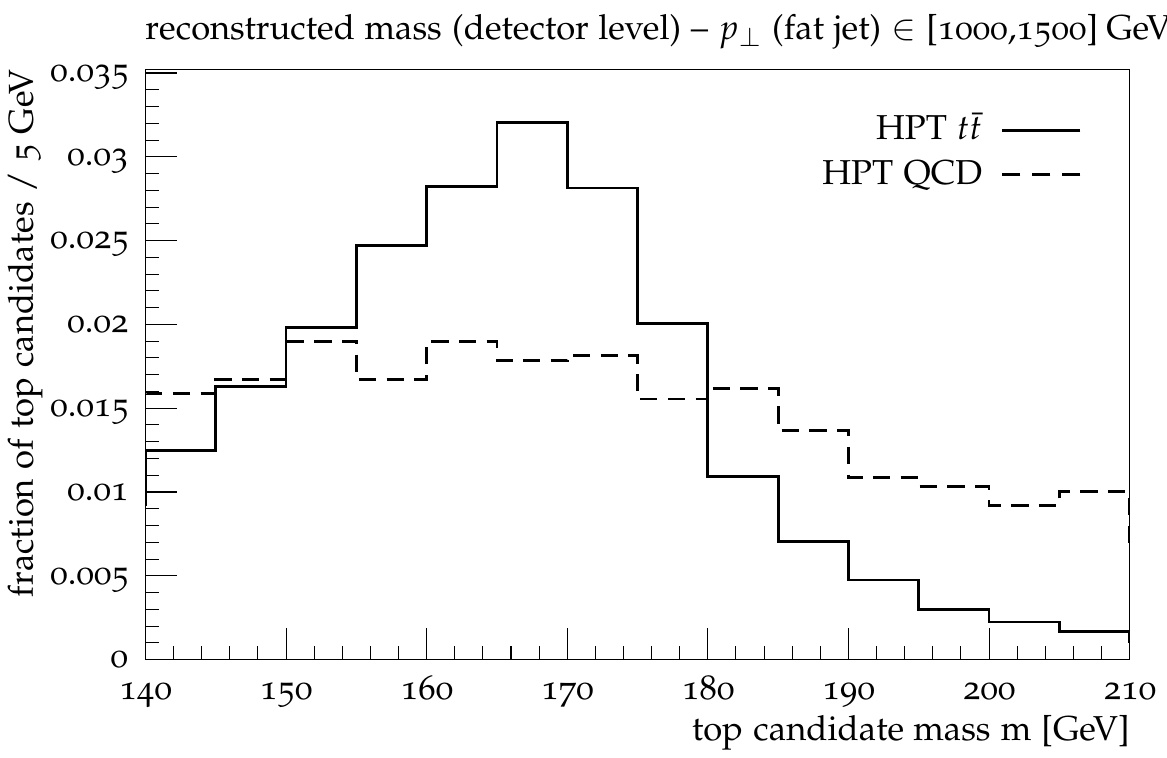}
  \includegraphics[width=0.45\textwidth]{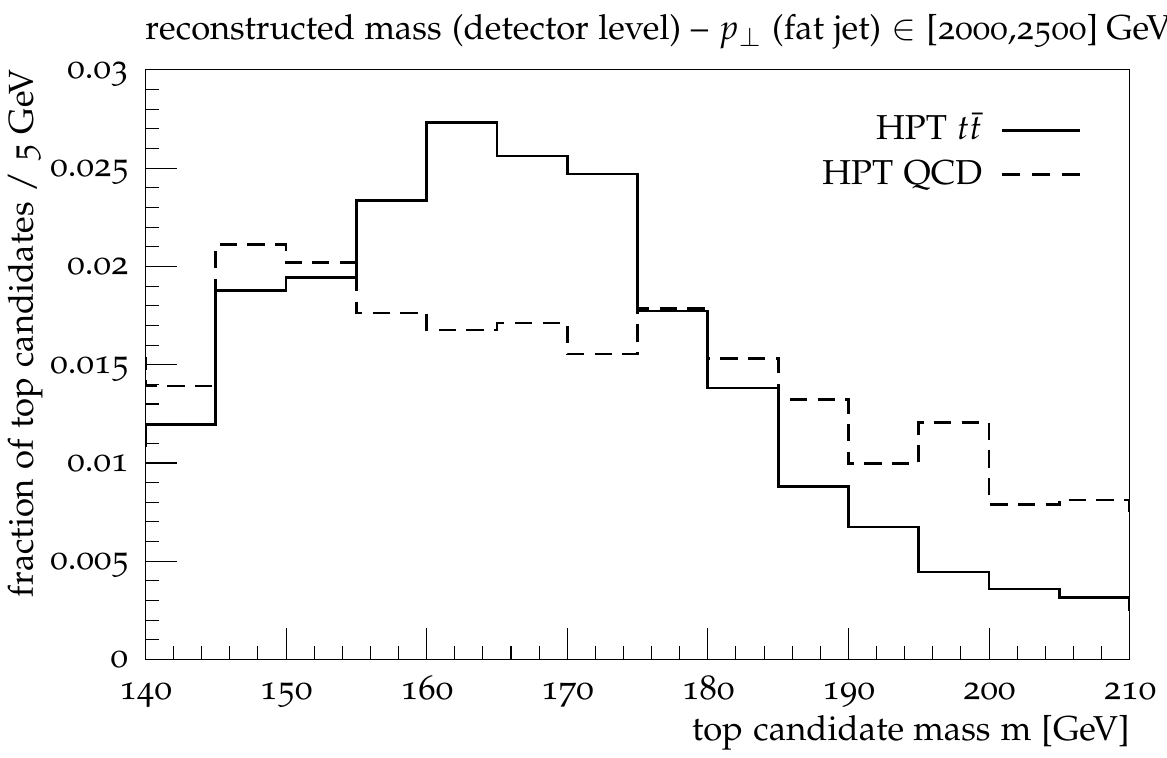}
 \end{center}
 \caption{Reconstructed top quark mass with the new \hptt at particle level (upper row) and detector level (\delphes) (lower row) for different transverse momentum ranges $p_{\perp,\mathrm{fatjet}}$.
 }
 \label{fig:t_reco_mass}
\end{figure}

\figref{t_reco_mass} shows the reconstructed top mass after applying the HPTTopTagger in two different transverse momentum windows, $p_{\perp,\mathrm{fatjet}} \in [1000,1500]\GeV$ (l.h.s.) and [2000, 2500]\GeV (r.h.s.). 
In the former, we generate background events with $\hat{p}_\perp\geq 700\GeV$ and in the latter with $\hat{p}_\perp\geq 1800\GeV$, to obtain a dropping $p_\perp$ distribution as expected from the SM process. 
The top mass distribution from the signal sample shows a clear peak at the correct top quark mass. While the detector simulation does not affect the position and width of the peak significantly, it is slightly sharper and more pronounced for $p_{\perp,\mathrm{fatjet}} \in [1000,1500]$~GeV compared to the window [2000, 2500]~GeV.

As track-based observables are not infrared safe \cite{chargedIR} non-perturbative contributions have to be taken into account to obtain well-defined results. For practical purposes, in event generators, hadronization models including fragmentation functions are used, which in turn follow perturbative evolution equations to obtain an extrapolated result at energies of interest from LEP data. Thus, track-based observables are subject to the parametrisation of non-perturbative physics. 

\begin{figure}[htb]
 \begin{center}
  \includegraphics[width=0.45\textwidth]{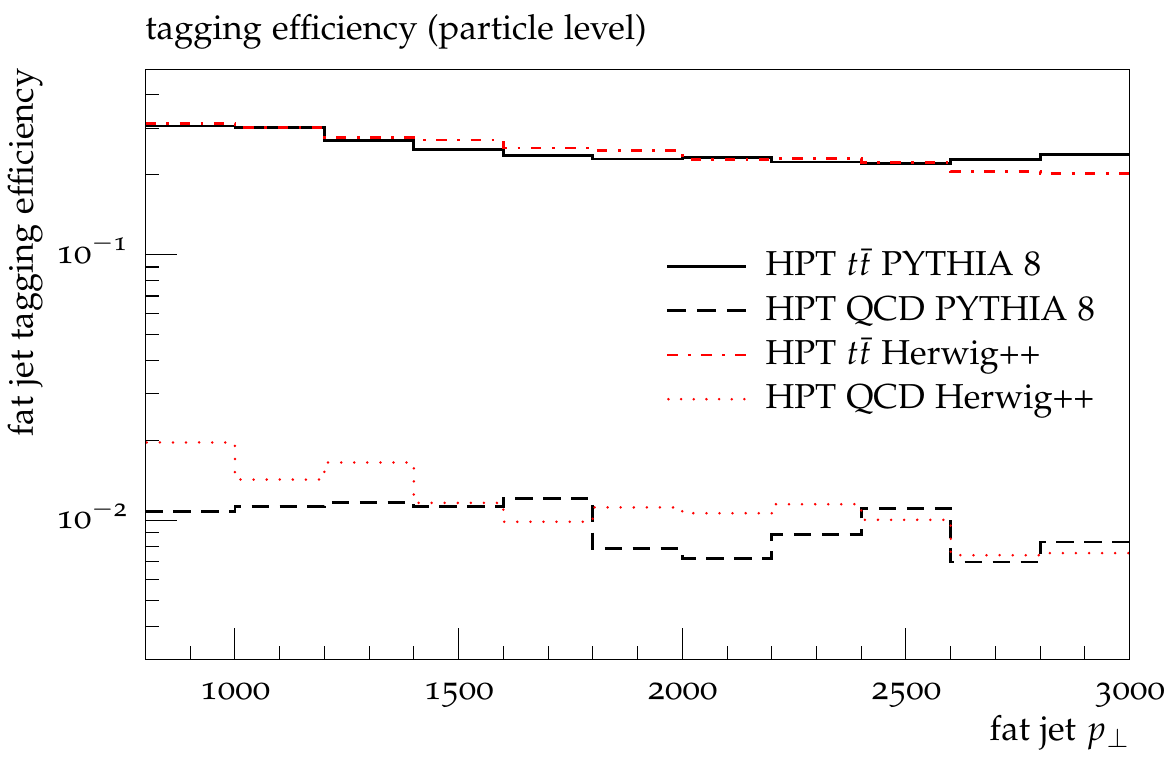}
  \includegraphics[width=0.45\textwidth]{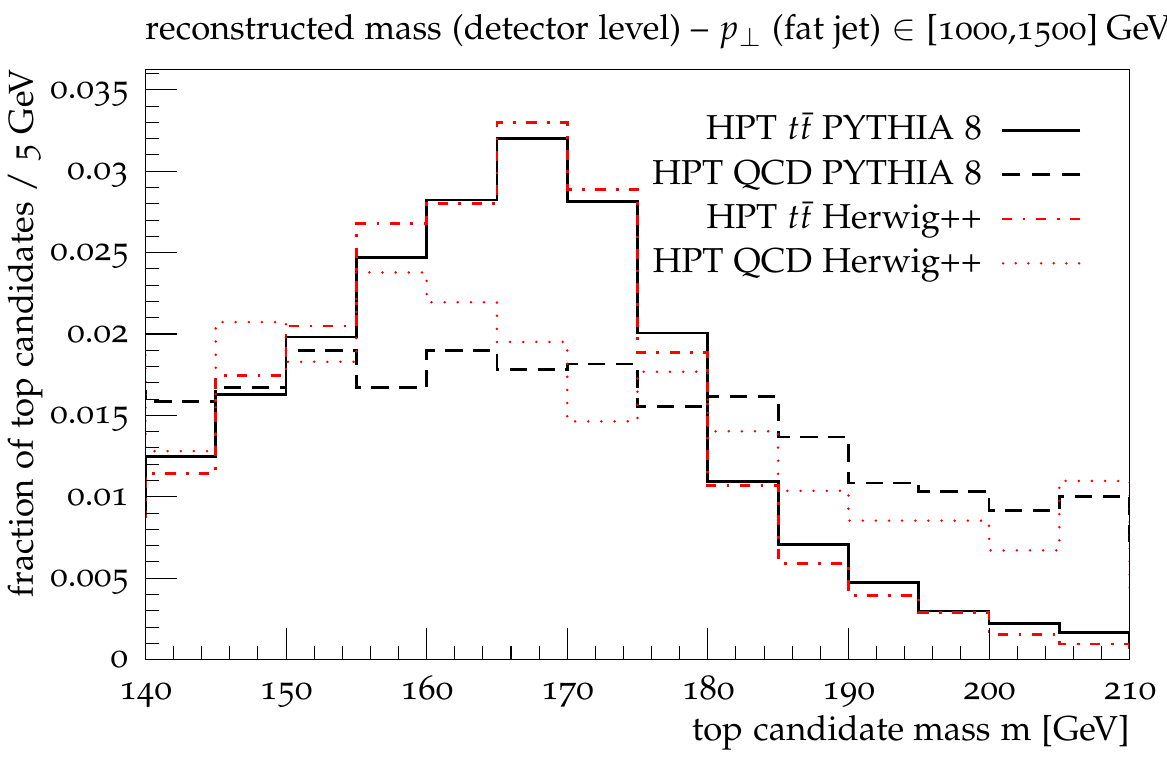}
 \end{center}
 \caption{Comparison of results for the new HPTTopTagger from samples generated with \pythia (black curves) and \herwigpp (red curves). Tagging efficiencies and reconstructed mass are shown at detector level for both top-initiated and QCD jets. 
}
 \label{fig:t_py_vs_hw}
\end{figure}

To estimate the impact different hadronisation models can have on the performance of the HPTTopTagger we compare the tagging efficiency and mass reconstruction for events generated with either \pythia or \herwigpp \cite{Bahr:2008pv}, see \figref{t_py_vs_hw}. We find that both event generators result in very similar signal efficiencies, i.e.~differences of $\mathcal{O}(10)\%$ at most. For the backgrounds the differences are slightly larger but still within generic uncertainties of event generators.

\subsection{Resonance search with highly boosted top quarks}
\label{sec:res:top}

The discovery of heavy resonances is a prime goal at the LHC and a possible future high-energy proton-proton collider. In many extensions of the Standard Model heavy resonances are predicted, leading to highly boosted top quarks. While for top quarks with intermediate boost focusing on (semi)leptonic top decays can lead to stronger exclusion limits \cite{ATLAS-CONF-2015-009}, at very high transverse momentum standard lepton-isolation requirements fail and the improved background rejection in leptonic final states ceases to compensate for the smaller branching ratio. Thus, we will focus on hadronic top decays only.

\subsubsection{Results for LHC}
\label{sec:res14}

\begin{figure}[htb]
 \begin{center}
  \includegraphics[width=0.45\textwidth]{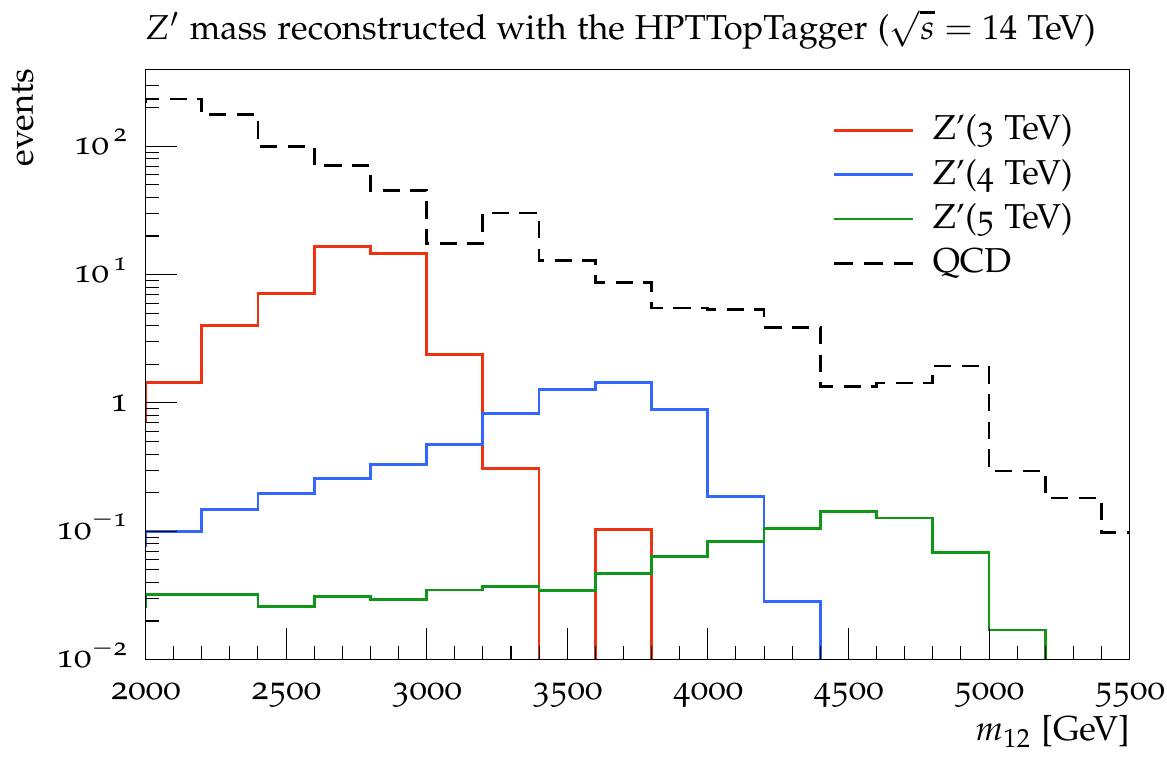}
    \includegraphics[width=0.45\textwidth]{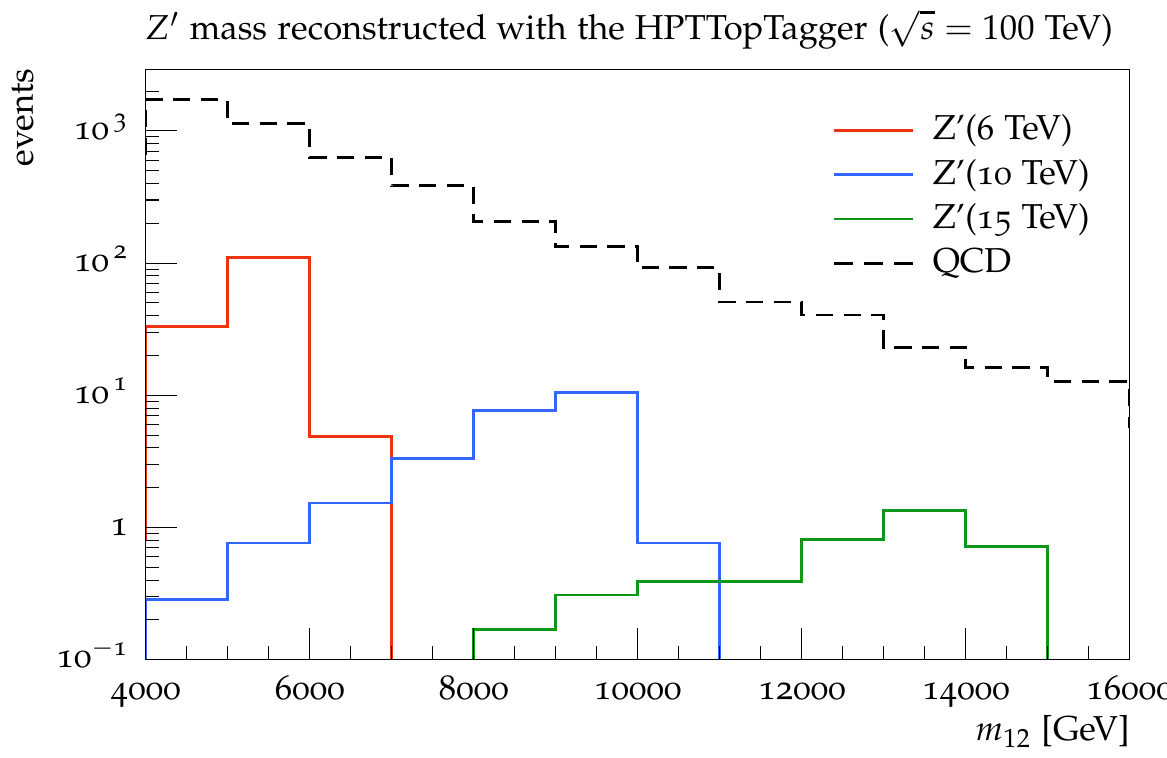}

 \end{center}
 \caption{Invariant mass of reconstructed top quarks after top tag $m_{12}$. The left figure corresponds to benchmark scenarios for $\sqrt{s}=14$ TeV and the right figure to $\sqrt{s}=100$ TeV. }
 \label{fig:toptagger_resonance}
\end{figure}

\figref{toptagger_resonance} shows expected event rates at the LHC with 14\TeV\  centre-of-mass energy and an integrated luminosity of 300~fb$^{-1}$.
We generate three signal benchmark scenarios with \pythia, choosing $m_{Z'}=(3,4,5)$ TeV with production cross-sections $\sigma_{Z'} = (3.45, 0.501,0.00962)$ fb and resonance widths $\Gamma_{Z'} = (95,128,160)$ GeV respectively.
We generate QCD dijet background events with $\hat{p}_\perp\geq 700$ GeV and find a total cross-section of 111.6~pb.

To reconstruct the $Z'$ resonances we require at least two fat jets clustered using the C/A jet algorithm with $R=0.8$. Input to the jet finder are all final state particles with $|\eta| < 4.9$, processed through \delphes. We further impose $p_{\perp,\mathrm{fatjet}} \geq 800$ GeV.

If both fat jets are tagged we reconstruct the resonance by summing the reconstructed top quarks' four-momenta $m^2_{12} = (p_{t_1} + p_{t_2})^2$, see \figref{toptagger_resonance}. While the resonance peak is clearly visible, it is shifted to values below the true resonance mass. In addition, for all resonance masses we find a long tail towards smaller $m_{12}$. The shift of the mass peak and the low-mass tail are more pronounced for heavier resonances. This is a known effect when reconstructing very heavy resonances. Top taggers aim to reconstruct top quarks that are close to being on-shell. However, top quarks produced in decays of heavy resonances have a large probability to radiate gluons before decaying. As these gluons are emitted in a wide angle around the top quark this radiation is lost when recombining the four-momenta of the reconstructed tops and their invariant mass is shifted towards smaller values.

We evaluate the required integrated luminosity to exclude the three benchmark resonances using a simple cut and count method. Based on the invariant mass distribution of the two reconstructed top quarks $m_{12}$ we choose the two bins with the best $S/B$ ratio and require $S/\sqrt{B} \simeq 2$.

\begin{table}
\begin{center}
\begin{tabular}{c|c|c|c|c|c}
 Resonance & $m_{12}$ window (TeV) & $S/B$ & $S/\sqrt{B}$ & $\sigma$ (fb) & $\sigma$ for $S/\sqrt{B}=2$ \\\hline
 $m_Z'=3$ TeV & 2.6-3.0 & 0.27 & 2.90 & 3.45 & 2.38 \\
 $m_Z'=4$ TeV & 3.4-4.0 & 0.13 & 0.69 & 0.501 & 1.45 \\
 $m_Z'=5$ TeV & 4.4-5.0 & 0.07 & 0.16 & 0.00962 & 0.12 
\end{tabular}
\end{center}
\caption{Results for search for $Z' \to t\bar{t}$ at the LHC14 in three benchmark scenarios. The last column shows the required production cross-section to achieve $S/\sqrt{B} = 2$ with $300~\mathrm{fb}^{-1}$. All numbers are based on the results provided in \figref{toptagger_resonance}.}
\label{tab:ZttLHC}
\end{table}

For our benchmark scenarios, with only 300~fb$^{-1}$ of data, a heavy $m_{Z'}=3\TeV$ can be excluded at 90\% CL.

Based on \tabref{ZttLHC} in \figref{toptagger_resonance} we show the necessary integrated luminosity for exclusion at 90\% CL, depending on the resonance's production cross-section.

\begin{figure}[htb]
 \begin{center}
  \includegraphics[trim=5cm 14cm 3cm 4cm, clip=true, width=0.49\textwidth]{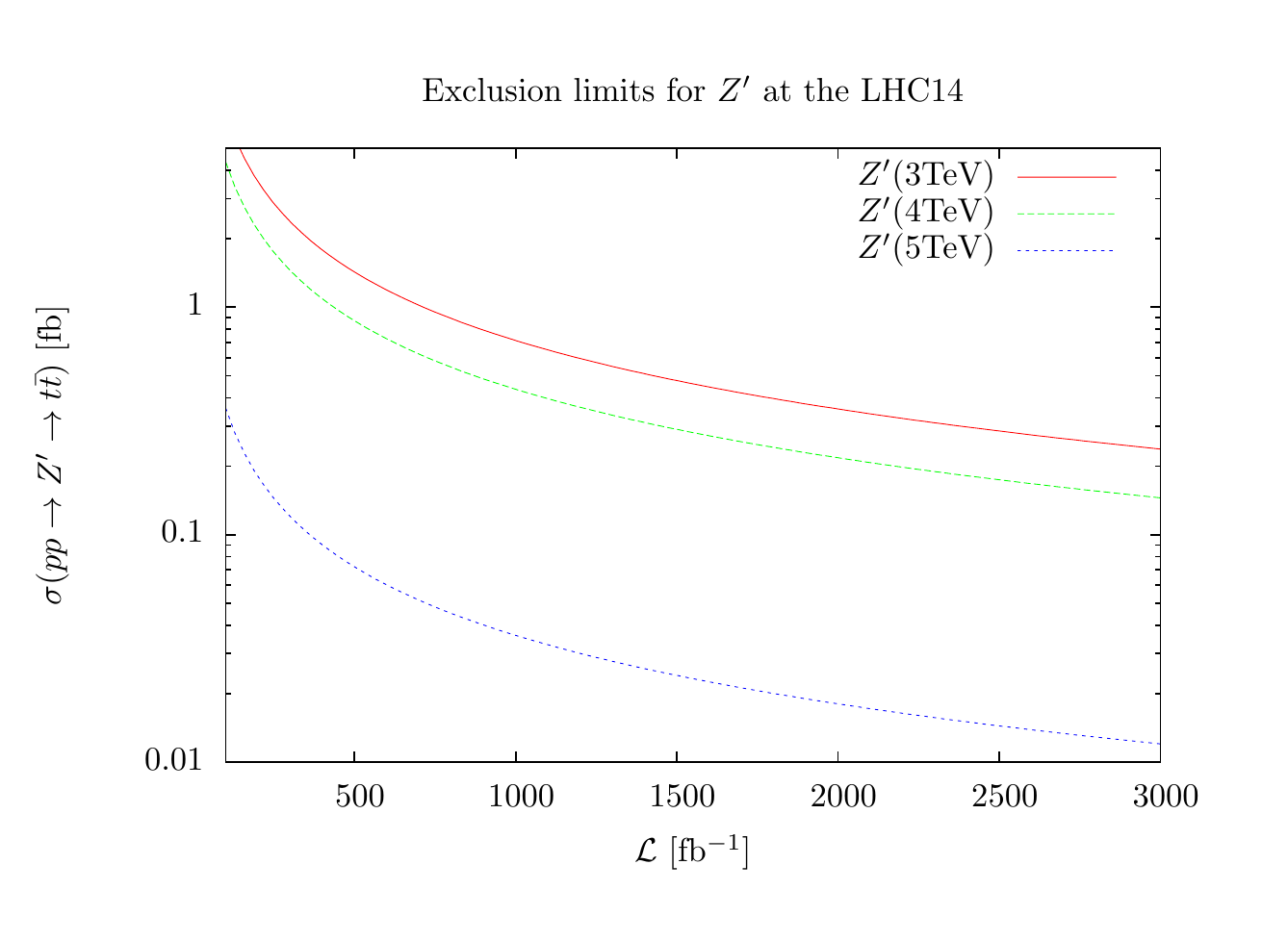}
  \includegraphics[trim=5cm 14cm 3cm 4cm, clip=true, width=0.49\textwidth]{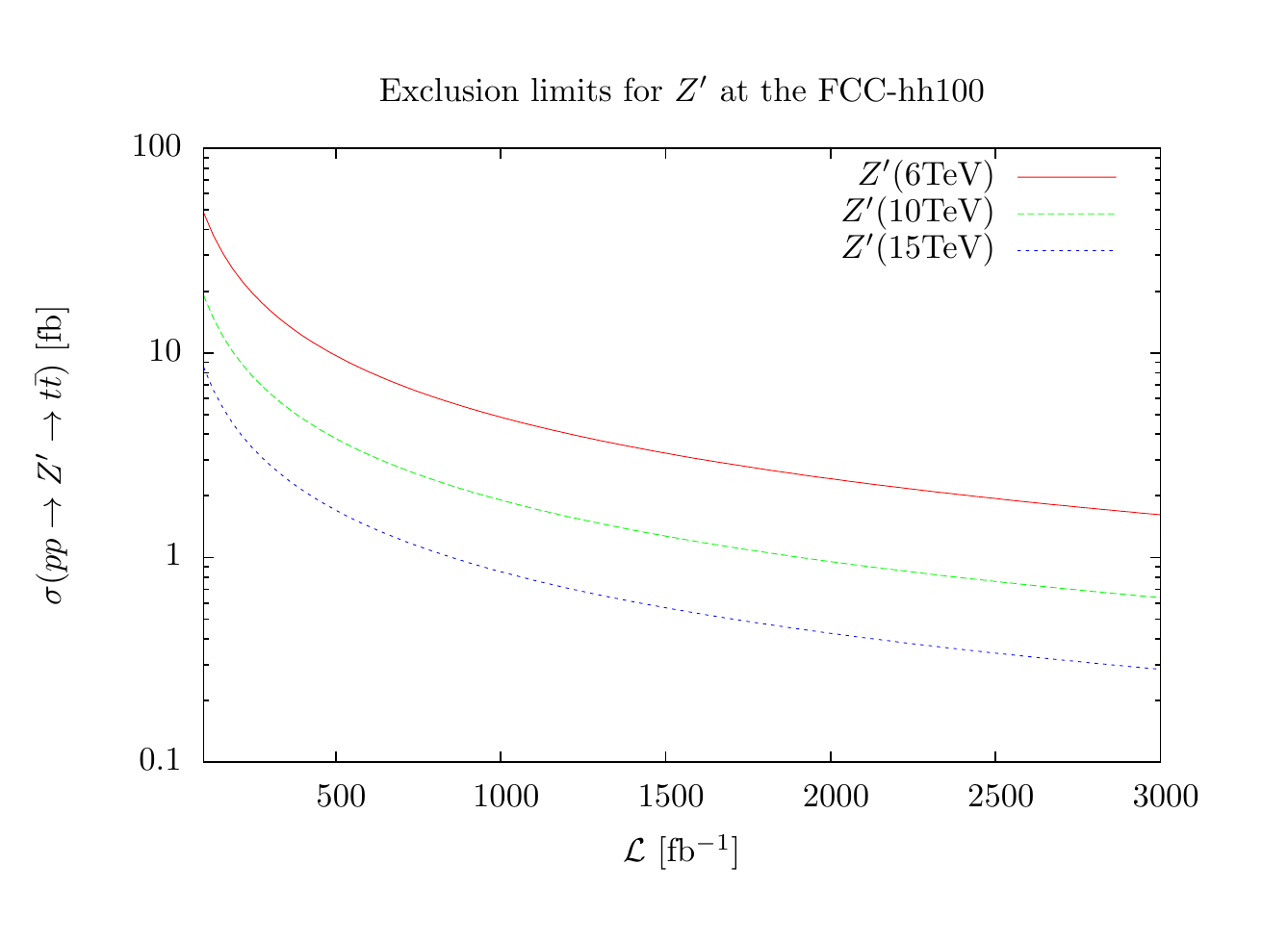}
 \end{center}
 \caption{Exclusion limits at 90\% CL for $Z'\to t\bar{t}$ with the new HPTTopTagger depending on production cross-section and integrated luminosity.
 The left figure corresponds to benchmark scenarios for $\sqrt{s}=14$ TeV and the right figure to $\sqrt{s}=100$ TeV.}
 \label{fig:toptagger_resonance_exclusion}
\end{figure}

\subsubsection{Results for FCC-hh 100 TeV}
\label{sec:res100}

We generate events as described in \secref{res14} but with a centre-of-mass energy of $\sqrt{s}=100\TeV$.
We employ the same analysis to reconstruct the heavy $Z'$ from two tagged tops, but now select fat jets with $p_{\perp,\mathrm{fatjet}} \geq 2000$ GeV.

Again, we generate three signal benchmark scenarios with \pythia, choosing $m_{Z'}=(6,10,15)$~TeV with production cross-sections $\sigma_{Z'} = (26.9, 0.986, 0.328)$ fb and resonance widths $\Gamma_{Z'} = (192,322,485)$~GeV respectively. 
We generate QCD dijet background events with $\hat{p}_\perp\geq 2000\GeV$ and find a total cross-section of 149.3~pb.

Expected event rates at 300\invfb are shown in \figref{toptagger_resonance} (r.h.s.) for all three benchmark masses and the QCD background.
The mass bins that give the largest significance are listed in \tabref{ZttFCC}.
Based on this table, we show 90\% CL exclusion limits as a function of the production cross-section and integrated luminosity in \figref{toptagger_resonance_exclusion} (r.h.s.).

\begin{table}[hb]
\begin{center}
\begin{tabular}{c|c|c|c|c|c}
 Resonance & $m_{12}$ window (TeV) & $S/B$ & $S/\sqrt{B}$ & $\sigma$(fb) & $\sigma$ for $S/\sqrt{B}=2$  \\\hline
 $m_Z'=6$ TeV & 5.0-6.0 & 0.098 & 3.29 & 26.9 & 16.39 \\
 $m_Z'=10$ TeV & 8.0-10.0 & 0.054 & 099 & 3.18 & 6.45\\
 $m_Z'=15$ TeV & 13.0-15.0 & 0.052 & 0.33 & 0.466 & 2.84
\end{tabular}
\end{center}
\caption{Results for search for $Z' \to t\bar{t}$ at the FCC-hh100 in three benchmark scenarios. The last column shows the required production cross-section to achieve $S/\sqrt{B} = 2$ with $300~\mathrm{fb}^{-1}$. All numbers are based on the results provided in \figref{toptagger_resonance}.}
\label{tab:ZttFCC}
\end{table}

\section{Tagging highly boosted gauge bosons}
\label{sec:wztag}

\subsection{The HPTWTagger and HPTZTagger algorithms}
\label{sec:gaugetag:algorithm}
Not only boosted top quarks are of interest for searches of new physics at the LHC and future colliders. Many extensions of the Standard Model predict heavy resonances decaying to $W$ or $Z$ bosons \cite{Wells:2008xg,Dugan:1984hq,Schmaltz:2005ky}. In addition, SM cross-sections for the emission of electroweak gauge bosons off quarks are strongly enhanced in energetic final states \cite{Bell:2010gi,Christiansen:2014kba,Krauss:2014yaa}. Thus, in this section we outline a proposal for a track-based tagger for highly boosted $W$ and $Z$ bosons.

In the hadronic decay of boosted electroweak gauge bosons, $W^\pm\to q\bar{q}'$ or $Z\to q\bar{q}$, the two daughter partons exhibit a typical separation $\Delta R\sim 2 m_{W / Z} / p_{\perp, W / Z}$.
A successful tagging algorithms has to resolve these subjets separately, where the finite resolution of the hadronic calorimeter becomes the bottleneck for large transverse momenta $p_\perp\gtrsim 500\GeV$.
Following the arguments above, it is straightforward to adapt the \hptt to $W$ and $Z$ boson decays. Due to the two-prong nature of the process, we replace the cut on mass ratios (``A-cut'') by a cut on the momentum fraction of the leading subjet,
\begin{align}
 f_{p_\perp} \equiv
 \frac{p_{\perp, j_1}}{p_{\perp, W/Z\text{ candidate}}} =
 \frac{p_{\perp, j_1}}{p_{\perp, j_1+j_2}} \,.
 \label{eq:fpt}
\end{align}
Fake candidates from QCD jets can generate a $W$- or $Z$-like mass via soft emissions and their $f_{p_\perp}$ is thus dominated by values close to 1.\\

We propose the following algorithm to reconstruct highly-boosted $W$ and $Z$ bosons:
\begin{enumerate}
 \item Define a fat jet $j$ using the C/A algorithm with $R=0.5$ and $p_\perp\geq500\GeV$.
 \item Discard all tracks that are not associated with $j$ or that have $p_\perp <500\MeV$.
 \item Scale the remaining track momenta as follows: Re-cluster $j$ with the anti-$k_T$ algorithm employing a small radius $R=0.2$, and calculate $\alpha_j\equiv E_\text{jet}/E_\text{tracks}$ for each subjet using its respective associated tracks. The momenta of those tracks are multiplied by $\alpha_j$.
 Combine the scaled tracks associated with the fat jet to a track-based jet $j_c$.
 \item Re-cluster $j_c$ using the anti-$k_T$ algorithm with $R=80\GeV/p_{\perp,j_c}$.
 If there are fewer than two subjets we consider the tag to have failed.
 \item Calculate the distance between the two leading subjets,
 $r\equiv \Delta R_{ij}$ and re-cluster $j_c$ with a new radius $R=0.8\, r$.
 If the combination of the new two leading subjets gives an invariant mass around the EW boson mass, $m_\text{candidate}\in m_{W/Z}\pm 10\GeV$, they form our boson candidate.
 \item If the momentum fraction of the two candidate subjets as defined in \eqref{fpt} satisfies
 $f_{p_\perp} \leq 0.85$ (for $W$ candidates) or $f_{p_\perp} \leq 0.80$ (for $Z$ candidates),
 we consider the EW boson tag successful.
\end{enumerate}

If higher tagging efficiencies are required, we suggest to modify steps 5 and 6 and allow a larger candidate mass window, $m_\text{candidate}\in m_{W/Z}\pm 15\GeV$, together with a looser cut on the subjet momentum fraction, $f_{p_\perp} \leq 0.85$. This setup is denoted ``working point 2'' (w.p.~2) below.

We are not concerned with discriminating between $W$ and $Z$ candidates.
While the mass peaks are fairly well separated (cf.~\figref{w_reco_mass}, \figref{z_reco_mass}), one may also consider additional distinguishing features such as (sub)jet charge \cite{Krohn:2012fg}.

\subsection{Performance}
\label{sec:gaugetag:performance}

We investigate the tagging efficiency and the reconstructed mass of the HPTWTagger as well as the HPTZTagger, as described in \secref{gaugetag:algorithm}.

The relevant signal fat jets for both taggers are taken from the production of a hypothetical charged heavy boson at the LHC, decaying into a $W$ and a $Z$ boson,
\begin{align}
 p p\to W'^\pm \to W^\pm Z \,.
\end{align}
We generate events with masses in the range $m_{W'}=3\cdots 6\TeV$.
We consider two semileptonic decay patterns,
$pp\to W'^\pm\to W^\pm Z\to (jj)(l^+l^-)$ to assess the HPTWTagger, and
$pp\to W'^\pm\to W^\pm Z\to (l^\pm\nu)(jj)$
for the HPTZTagger, respectively.
Background events with a generator-level cut $\hat{p}_\perp=400\GeV \cdots 2500\GeV$ we generate from $pp\to Zj\to (l^+l^-)j$ for the HPTWTagger.
The background process to assess the HPTZTagger is $pp\to W^\pm j\to (l^\pm\nu)j$ with the same generator-level cuts.
We generate all events with \pythia~8~\cite{Sjostrand:2007gs} at centre-of-mass energy $\sqrt{s}=14\TeV$.
Fat jets are clustered from stable particles with pseudorapidity $|\eta|<4.9$ using the C/A jet algorithm with resolution parameter $R=0.5$ and $p_\perp\geq 500$ GeV. To assess the signal efficiency we match the fat jet to MC truth $W$ / $Z$ bosons by requiring $\Delta R(j,\mathrm{W/Z}) < 0.4$ and select jets with $|\eta_j| < 2.5$.
Additionally, we require the fat jet to be isolated from the MC truth leptons from the other gauge boson.
For a leptonically decaying $Z$ boson, the condition is $\Delta R(j,\mathrm{lepton}) > 0.6$.
The isolation criterion for a leptonic $W$ boson decay reads
$\Delta R(j,\mathrm{lepton / neutrino})>0.6$.

\begin{figure}[htb]
 \begin{center}
  \includegraphics[width=0.45\textwidth]{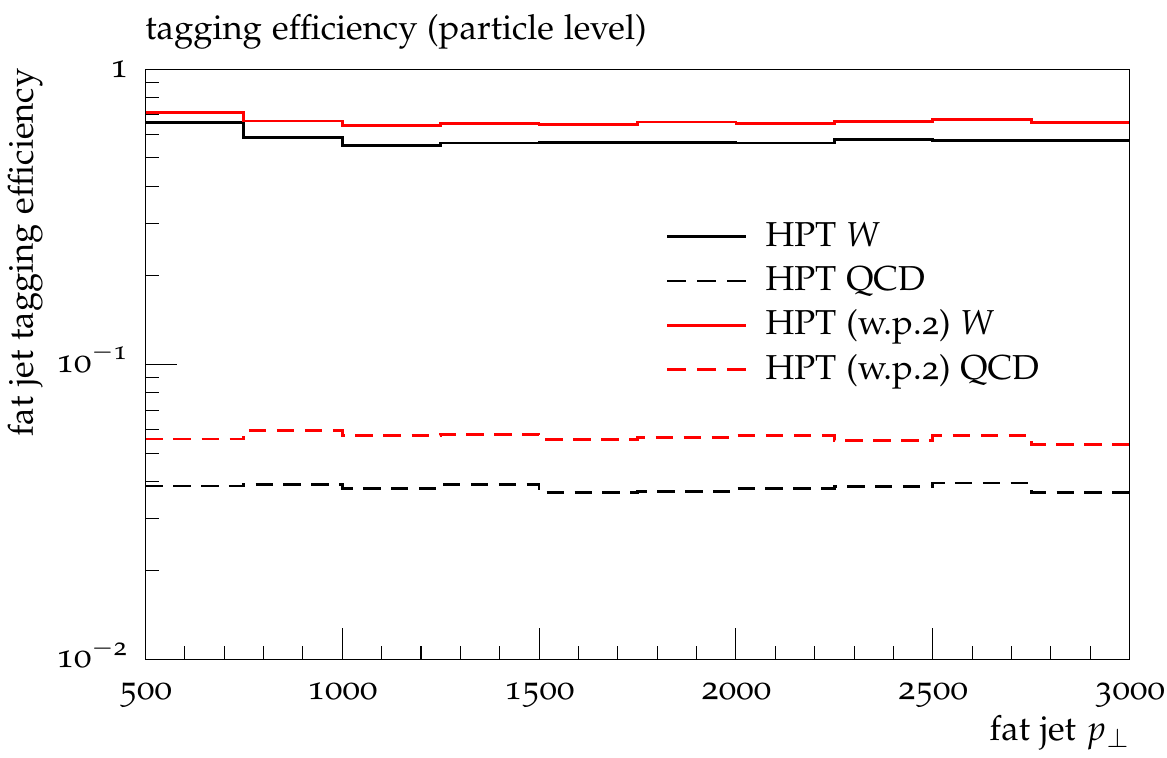}
  \includegraphics[width=0.45\textwidth]{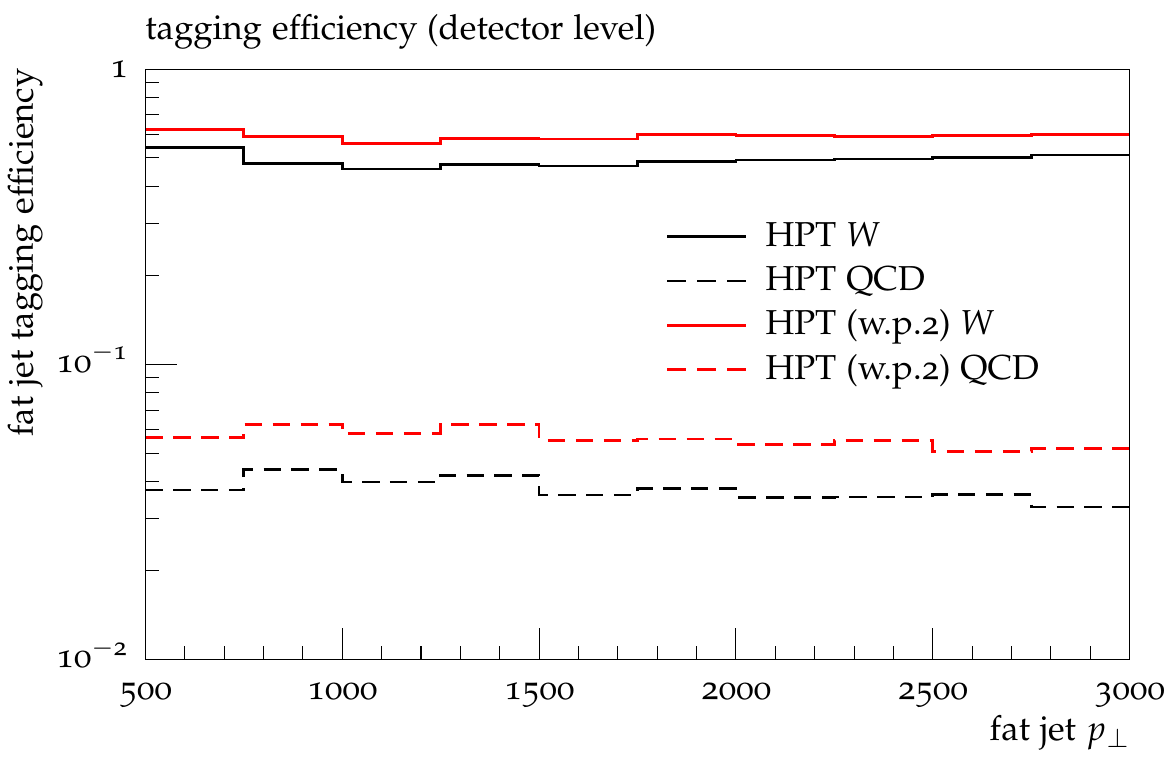}
 \end{center}
 \caption{Tagging efficiency of the HPTWTagger at particle level (l.h.s.) and detector level (\delphes) (r.h.s.) for boosted $W$ bosons (solid lines). Standard Model QCD background mistag rates are given by the dashed lines. Results are shown for the standard setup (black) and the higher-efficiency working point 2 (red). 
 }
 \label{fig:w_efficiency}
\end{figure}

\figref{w_efficiency} shows the efficiencies for tagging highly boosted $W$ bosons with the HPTWTagger and background mistag rates for both working points suggested in \secref{gaugetag:algorithm}.
Results are presented at particle level (left) and detector level after running \delphes (right).
We use all event samples as listed above to generate the plots in order to achieve good statistics over the whole $p_\perp$ range.

\begin{figure}[htb]
 \begin{center}
  \includegraphics[width=0.45\textwidth]{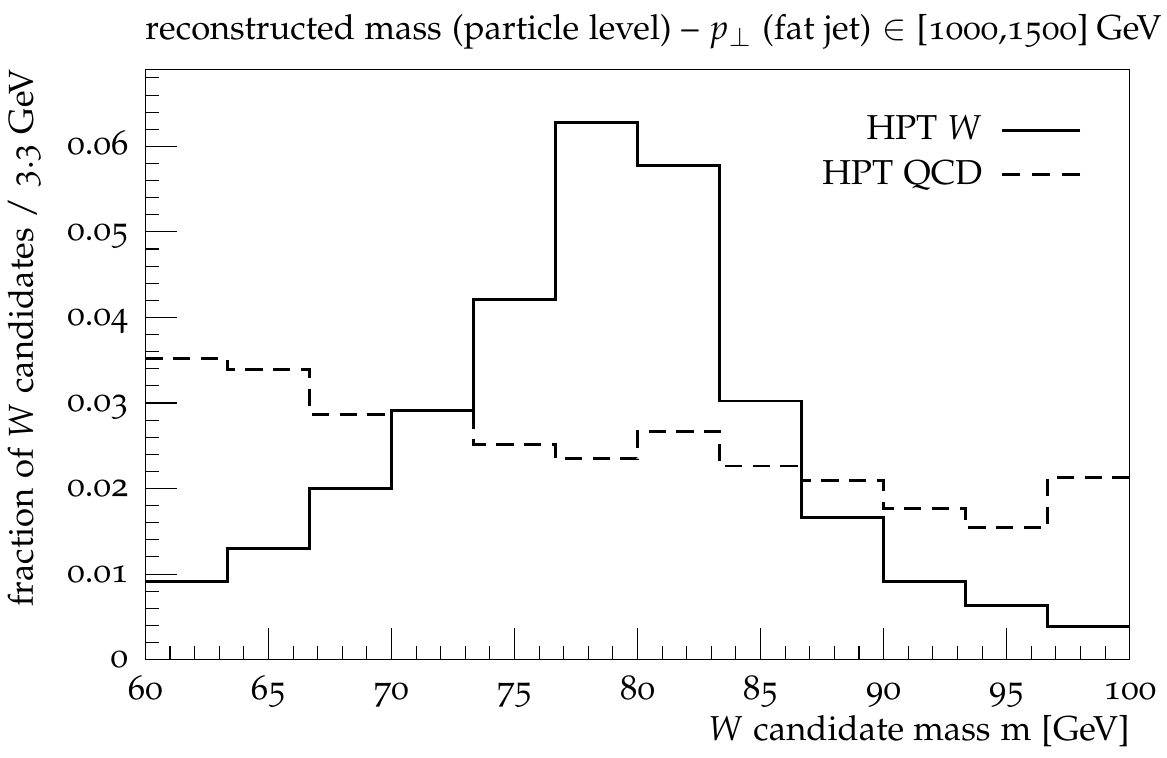}
  \includegraphics[width=0.45\textwidth]{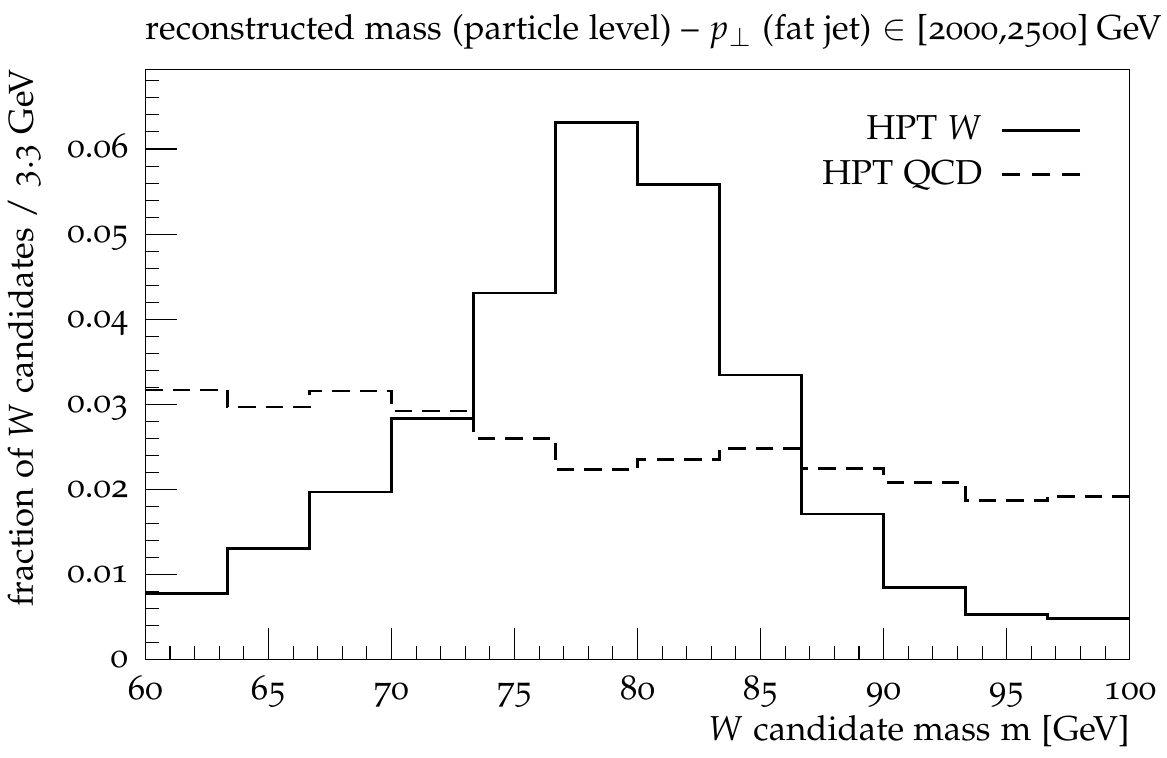}\\
  \includegraphics[width=0.45\textwidth]{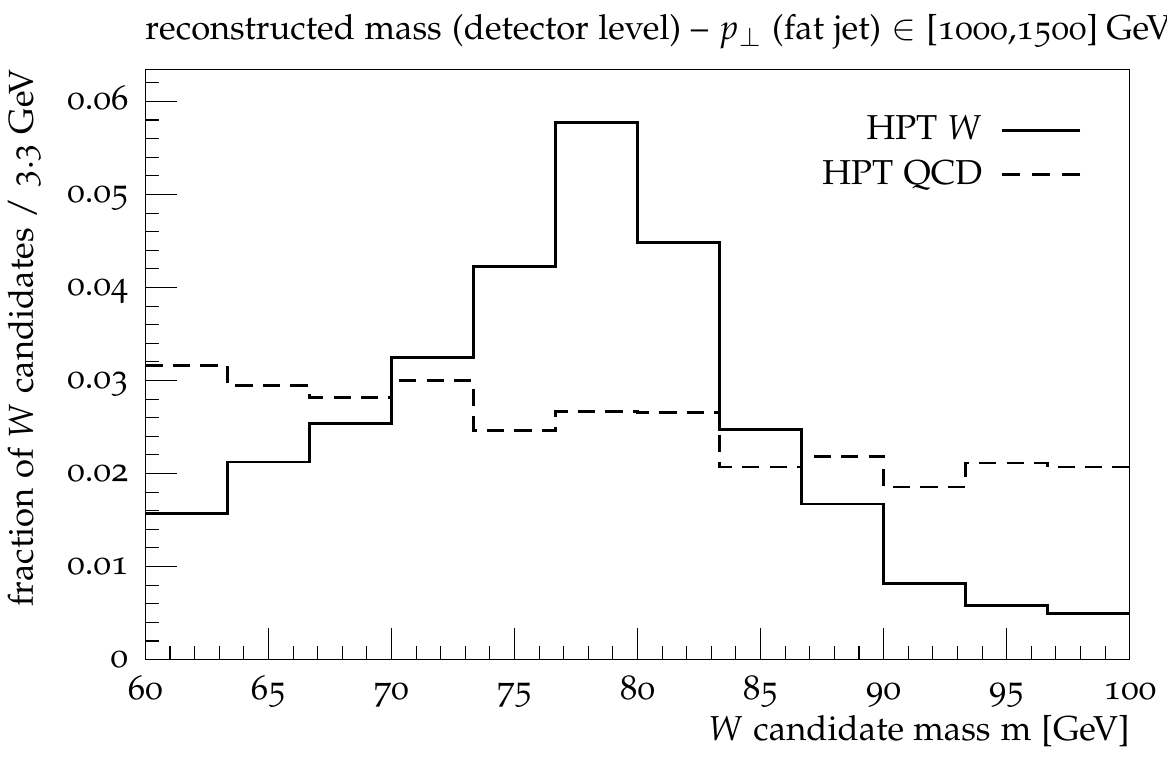}
  \includegraphics[width=0.45\textwidth]{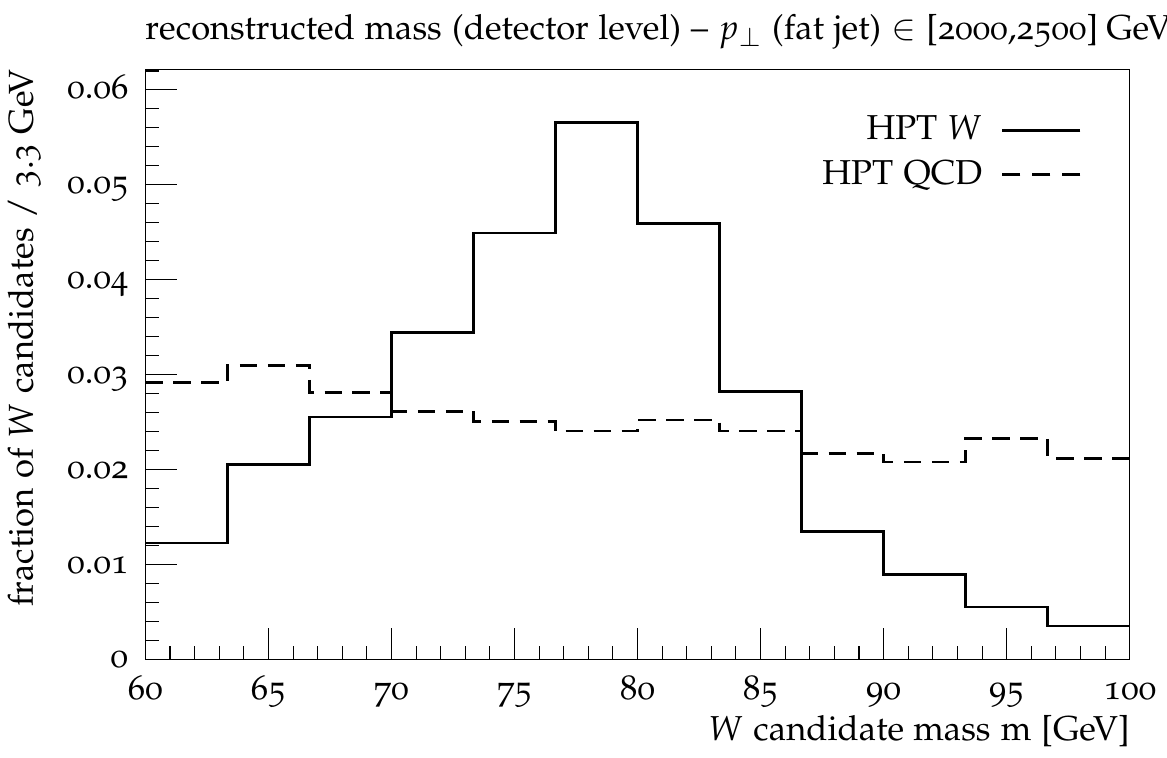}
 \end{center}
 \caption{Reconstructed $W$ boson mass with the HPTWTagger at particle level (upper row) and detector level (\delphes) (lower row) for different $p_{\perp,\mathrm{fatjet}}$. 
 }
 \label{fig:w_reco_mass}
\end{figure}

We show the reconstructed $W$ boson mass in~\figref{w_reco_mass} in two different $p_\perp$ bins, also both at particle and detector level.
In the $p_\perp\in [1000,1500]\GeV$ bin (left), we only consider the $m_{W'}=3\TeV$ sample for signal events. In order to obtain a dropping $p_\perp$ distribution as expected from the SM process, we generate background events with fixed generator-level cut $\hat{p}_\perp\geq 1000\GeV$ .
At higher transverse momenta in the $p_\perp\in [2000,2500]\GeV$ bin (right), we simulate signal events with $m_{W'}=5\TeV$ and for the background we impose a generator-level $\hat{p}_\perp\geq 2000\GeV$.

\begin{figure}[htb]
 \begin{center}
  \includegraphics[width=0.6\textwidth]{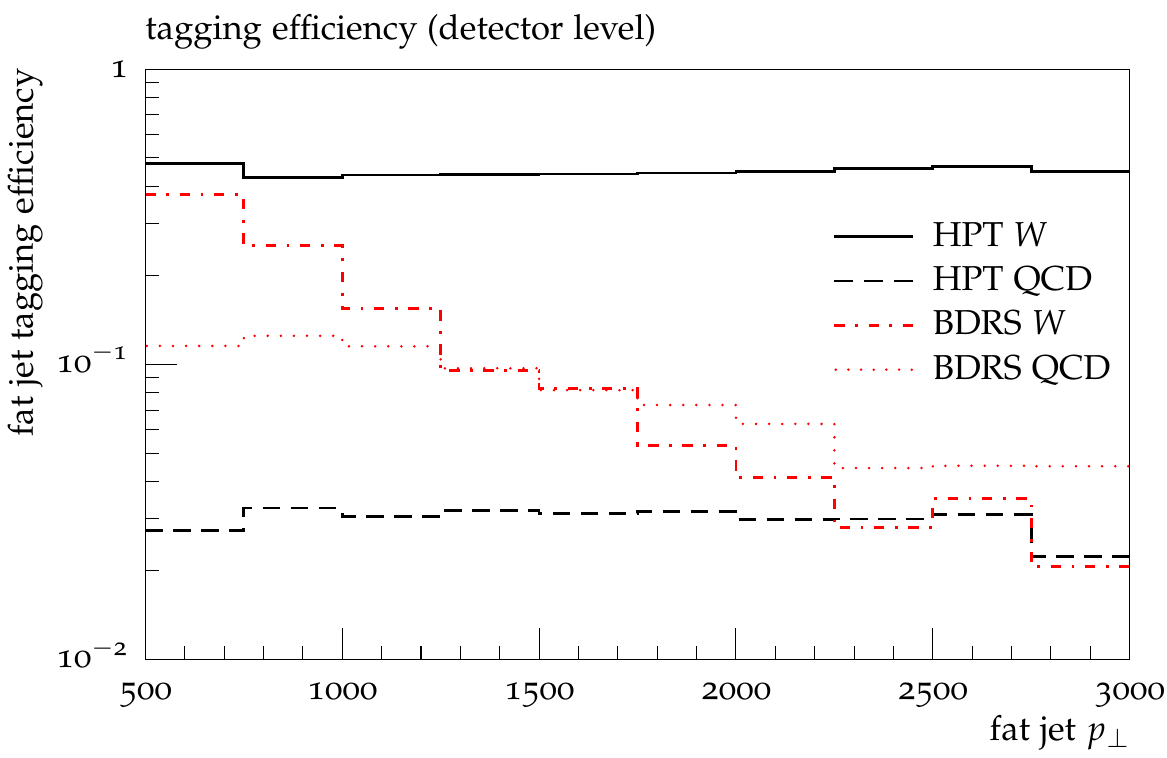}
 \end{center}
 \caption{Comparison of the tagging efficiencies of the HPTWTagger (black lines) and a BDRS-like tagger (red lines) at detector level (\delphes) for boosted $W$ bosons. Standard Model QCD background mistag rates are also given.  }
 \label{fig:w_efficiency_vs_bdrs}
\end{figure}

In~\figref{w_efficiency_vs_bdrs} we compare the tagging efficiency of our HPTWTagger to a BDRS-like algorithm~\cite{bdrs}.
The latter sequentially unclusters the fat jet, discarding the less heavy subjet at each step until a mass drop is found,
\begin{align}
\label{eq:bdrs1}
 \frac{\max{m_{j_i}}}{m_j}<0.67 \quad,\qquad
 \frac{\min (p_{\perp,j_i}^2) \Delta R_{j_1j_2}^2}{m_j^2} \sim \frac{\min{p_{\perp,j_i}}}{\max{p_{\perp,j_i}}} > 0.09 \,.
\end{align}
If no mass drop is found, the candidate is discarded.
The two subjets are then filtered with a radius
\begin{align}
\label{eq:bdrs2}
 R_\text{filtering} = \max\left( 0.2, \min(0.3, \Delta R_{j_1j_2}/2) \right)
\end{align}
and the three hardest fitered subjets are summed to form the $W$ boson candidate.
In accordance with out algorithm, we consider the tag successful if the invariant mass lies within $m_W\pm 10\GeV$.

For the performance of the BDRS algorithm as implemented according to Equations~(\ref{eq:bdrs1}) and (\ref{eq:bdrs2}) quickly deteriorates for increasing jet $p_\perp$, see \figref{w_efficiency_vs_bdrs}. For $p_{\perp,j} > 1500$ GeV it is more likely to obtain a $W$-tag by with a QCD jet than with a $W$ decay. While optimising the parameters entering \eqref{bdrs1} might help to recover efficiency, the minimal angular separation of calibrated filtered subjets $\Delta R_{j1,j2} > 0.2$ eventually limits the applicability of this tagger for highly boosted resonances.

\begin{figure}[htb]
 \begin{center}
  \includegraphics[width=0.45\textwidth]{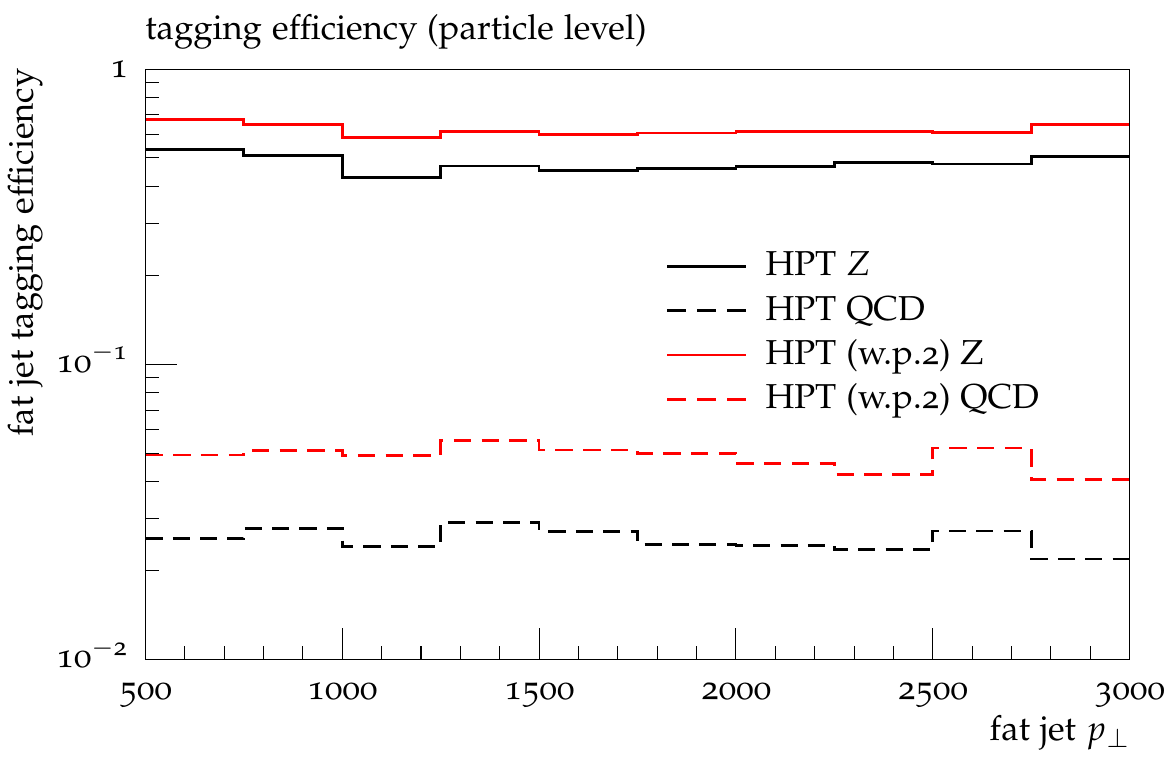}
  \includegraphics[width=0.45\textwidth]{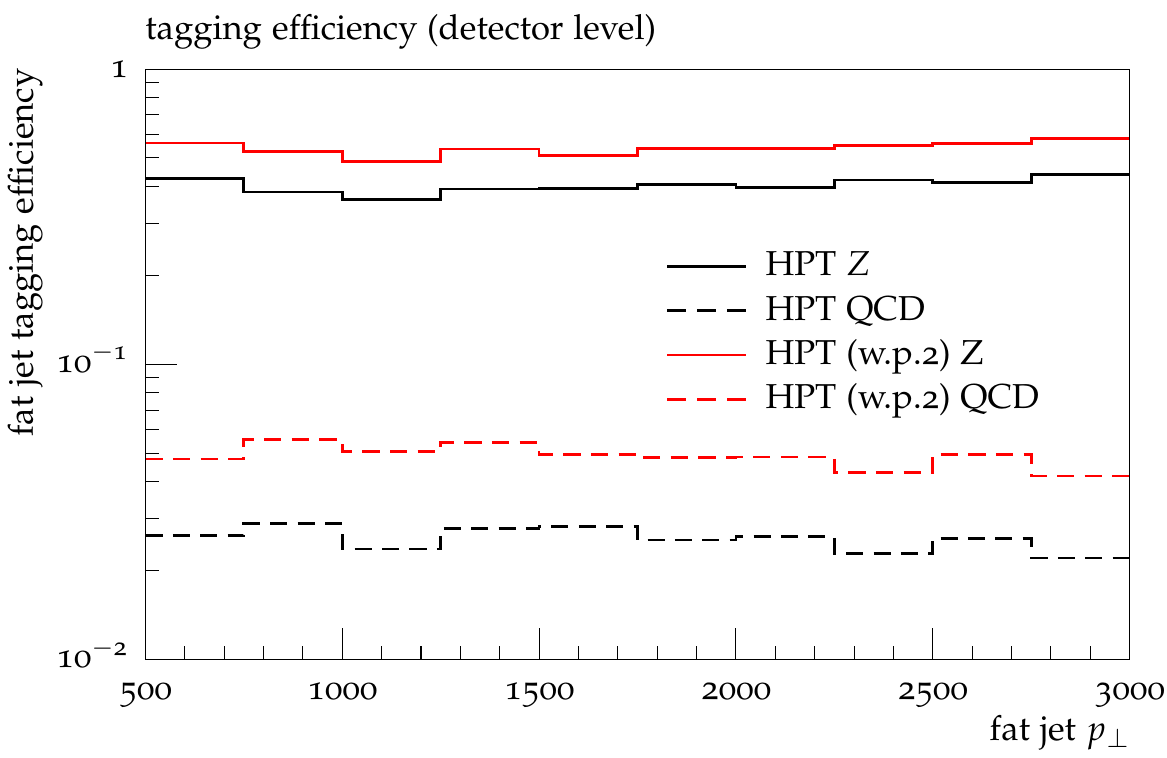}
 \end{center}
 \caption{Tagging efficiency of the HPTZTagger at particle level (l.h.s.) and detector level (\delphes) (r.h.s.) for boosted $Z$ bosons (solid lines). Standard Model QCD background mistag rates are given by the dashed lines. Results are shown for the standard setup (black) and the higher-efficiency working point 2 (red).
 }
 \label{fig:z_efficiency}
\end{figure}

We show the boosted $Z$ boson tagging efficiencies of the HPTZTagger in~\figref{z_efficiency} for both working points and both at particle level (left) and detector level (right).
Again, tagging efficiency and background rejection are stable over a large range of transverse momentum from $p_\perp=500\GeV$ up to 3\TeV.
Throughout the remainder of this section, the cuts imposed at generator level, i.e.~signal $m_{W'}$ and $\hat{p}_\perp$ for the background,  are the same as in the corresponding plots of the HPTWTagger.

\begin{figure}[htb]
 \begin{center}
  \includegraphics[width=0.45\textwidth]{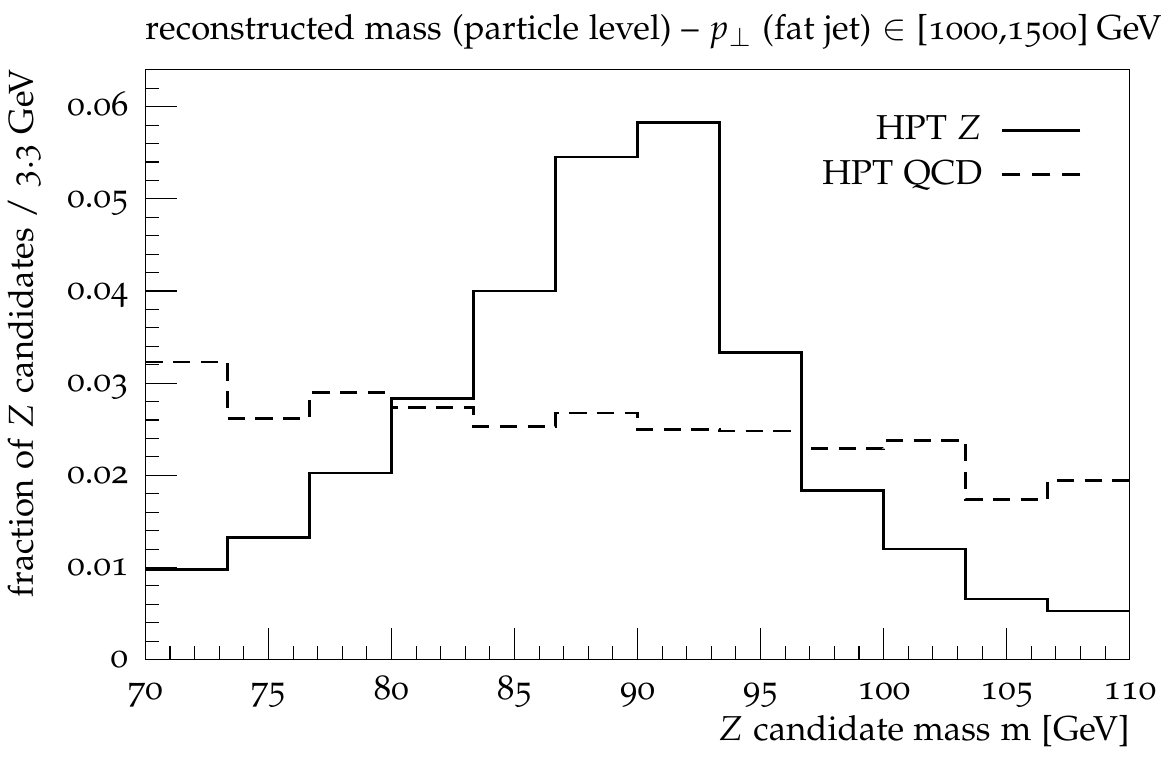}
  \includegraphics[width=0.45\textwidth]{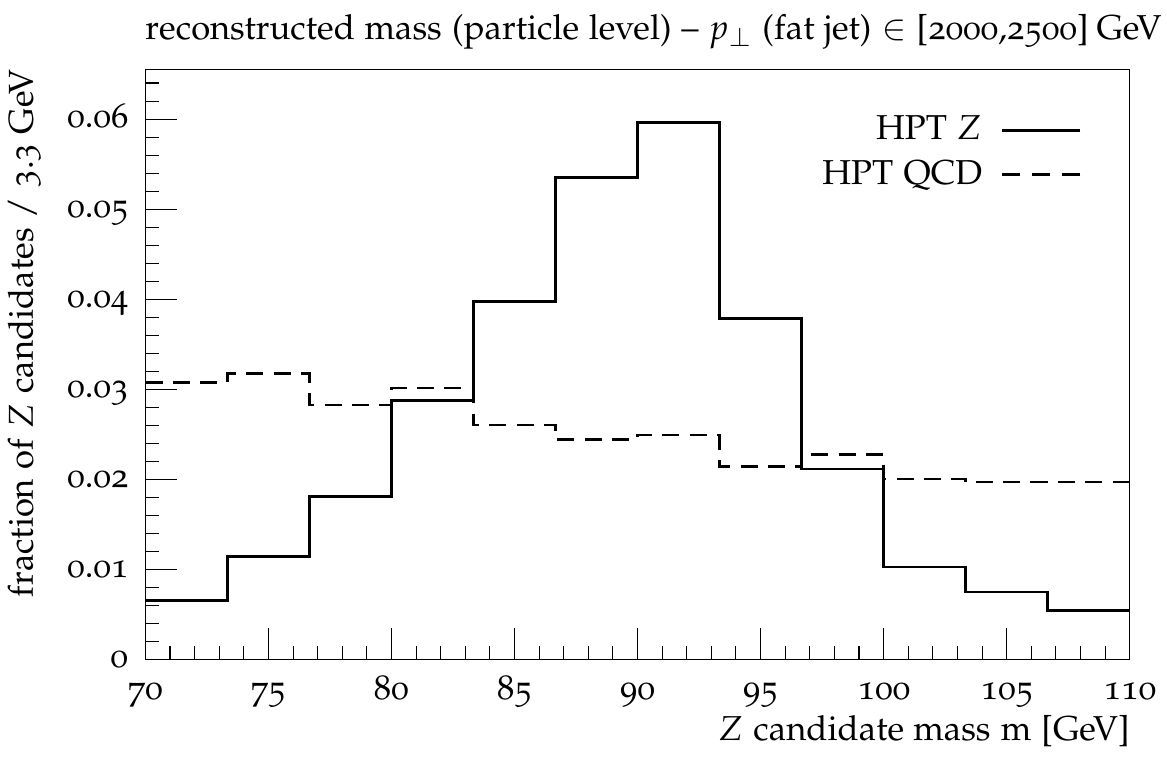}\\
  \includegraphics[width=0.45\textwidth]{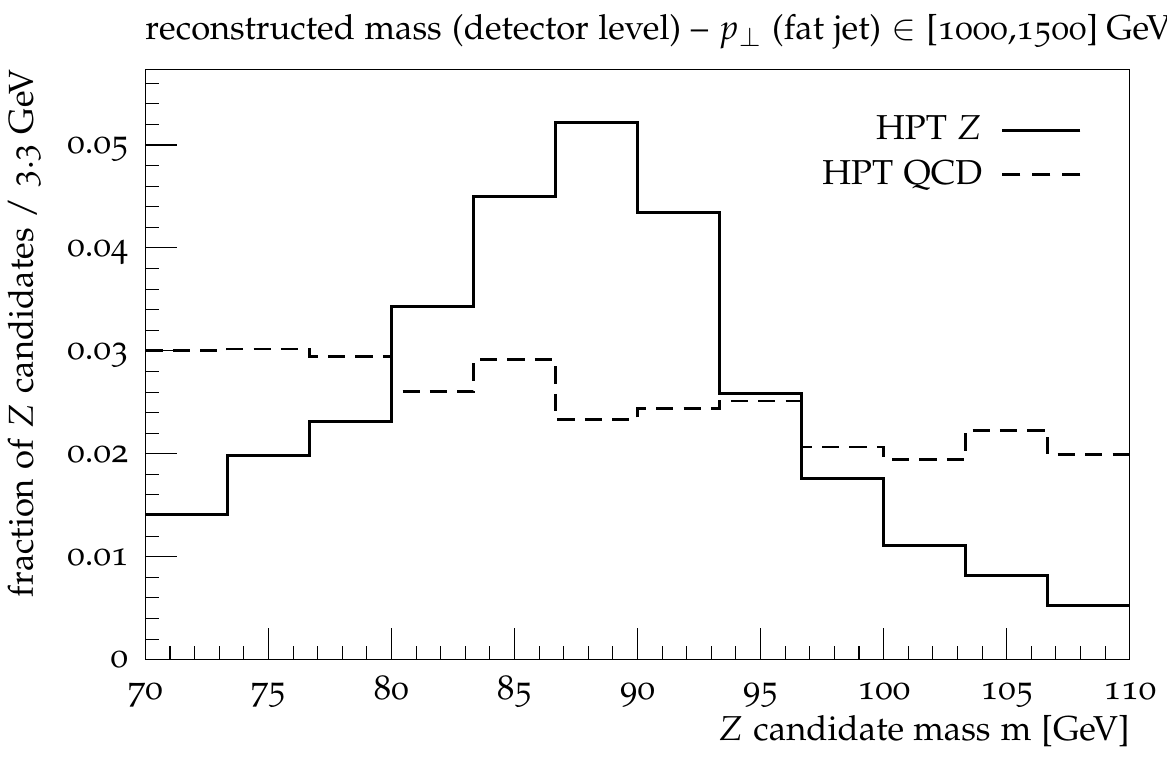}
  \includegraphics[width=0.45\textwidth]{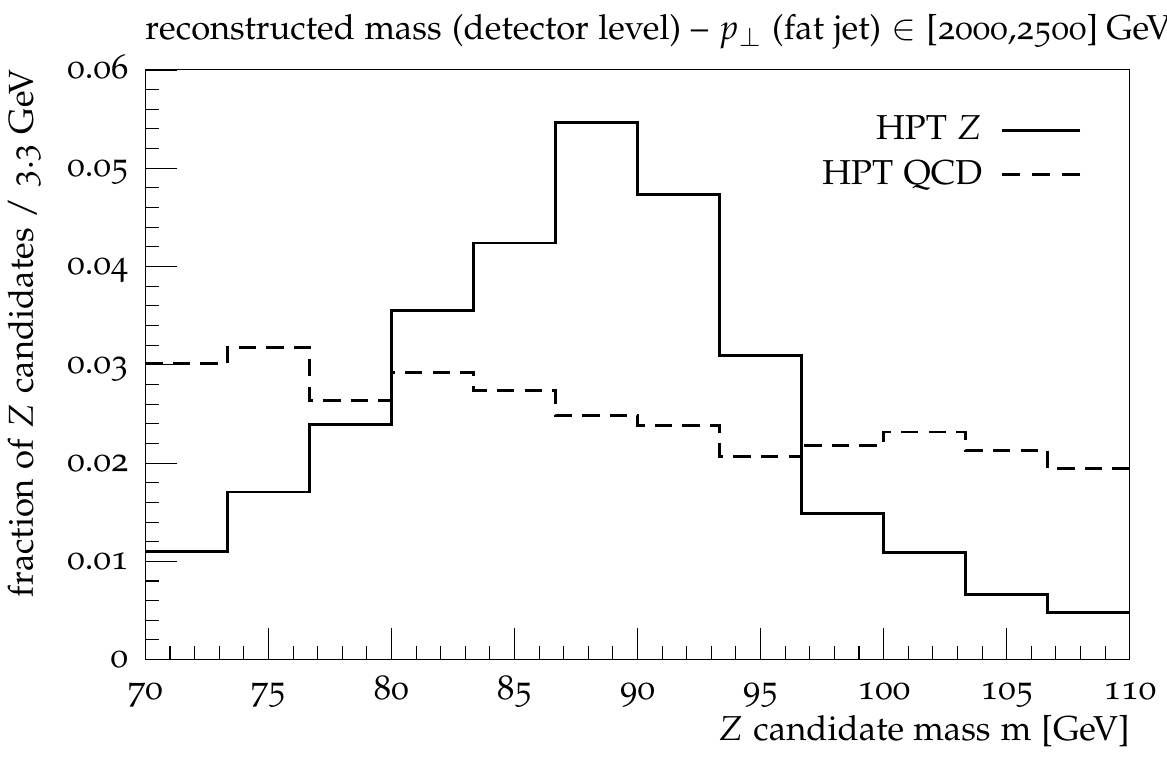}
 \end{center}
 \caption{Reconstructed $Z$ boson mass with the HPTZTagger at particle level (upper row) and detector level (\delphes) (lower row) for different $p_{\perp,\mathrm{fatjet}}$.
 }
 \label{fig:z_reco_mass}
\end{figure}

\figref{z_reco_mass} depicts the reconstructed mass of the boosted $Z$ boson.
Both in the highly boosted (left) and very highly boosted regimes (right), a peak at the MC truth mass is found over a flat background mass distribution.
We consider signal events with $m_{W'}=3\TeV$ (5\TeV) and background events with $\hat{p}\geq 1000\GeV$ (2000\GeV) in the $p_\perp\in [1000,1500]\GeV$ ([2000,2500]\GeV) bins to approximate realistic $p_\perp$ distributions.
No degredation is found in the reconstructed mass when going to high fat jet transverse momentum.

\subsection{Resonance search with highly boosted gauge bosons}
\label{sec:res:gauge}

Production of a charged heavy boson,
$p p\to W'^\pm \to W^\pm Z$,
as discussed in \secref{gaugetag:performance} is ideally suited for resonance searches at the LHC.
We investigate $W'$ masses in the range $m_{W'}=3\cdots 6\TeV$.
Again we consider two semileptonic scenarios, depending on whether the $W$ boson or the $Z$ boson decays hadronically.
In either case, a fat jet is formed with the C/A algorithm with parameters $R=0.5$ and $p_\perp\geq 800\GeV$.
The other gauge boson we force to decay leptonically and we assume perfect reconstruction of the visible electrons and muons.

In the first scenario $pp\to W'^\pm\to W^\pm Z\to (jj)(l^+l^-)$, we apply the HPTWTagger to tag and reconstruct the $W$ boson.
Identification of the decay of the $Z$ boson into a pair of charged leptons ($e^+e^-$, $\mu^+\mu^-$) reduces the relevant Standard Model backgrounds to $p p\to Z j$.
Here we assume perfect reconstruction of the leptons from the $Z$ boson decay.
A fat jet we reject if close to any of the leptons from the $Z$ decay,
$\Delta R(j,\mathrm{lepton}) < 0.6$.
We generate background events in bins of generator-level $\hat{p}_\perp$ ranging from $700\GeV$ to $2.5\TeV$, and $\hat{p}_\perp\geq 2.5\TeV$.

\begin{figure}[htb]
 \begin{center}
  \includegraphics[width=0.45\textwidth]{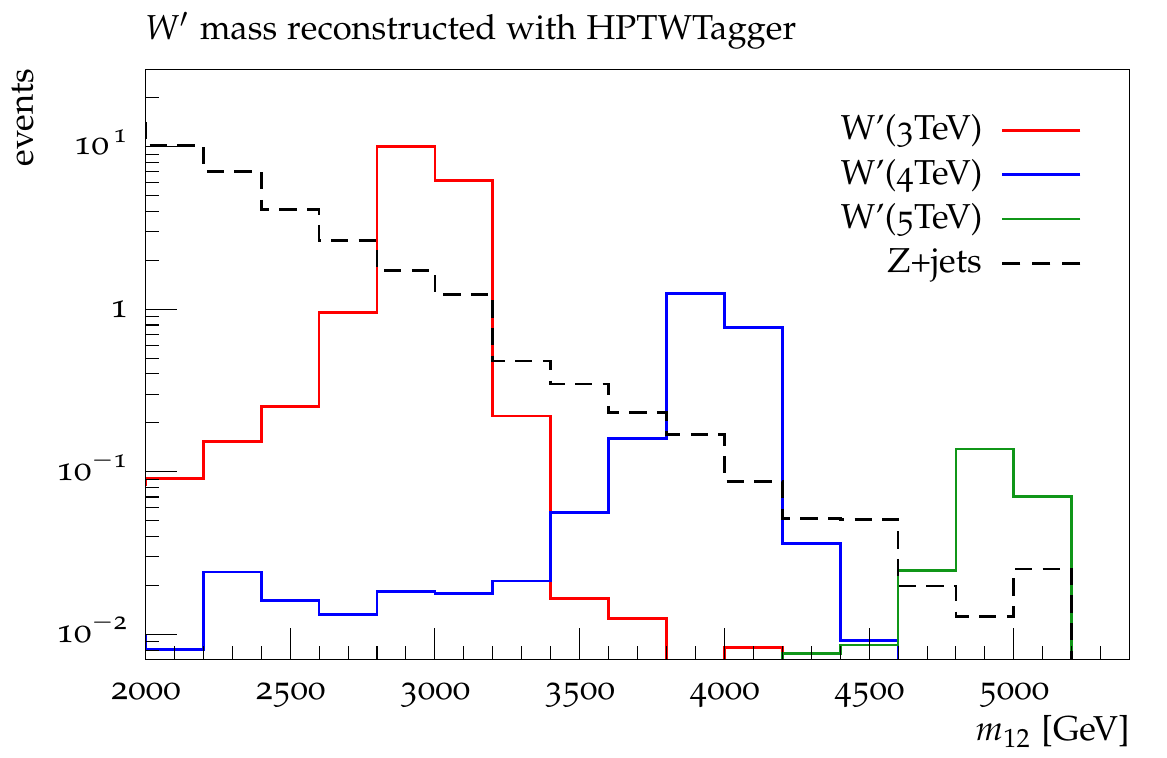}
  \includegraphics[width=0.45\textwidth]{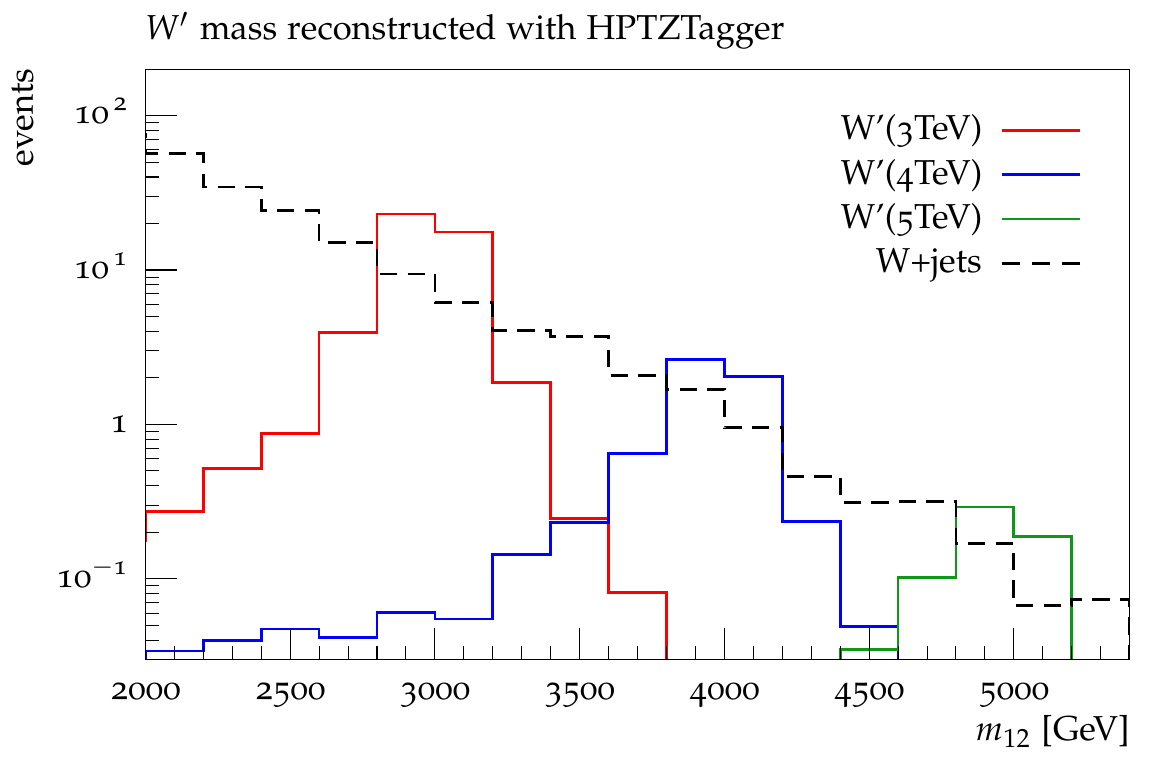}
 \end{center}
 \caption{Invariant mass of a heavy boson $W'^\pm\to W^\pm Z$ for three different resonance masses $m_{W'}=3\TeV$ (red), $m_{W'}=4\TeV$ (blue), and $m_{W'}=5\TeV$ (green).
 In the left panel the hadronically decaying $W$ boson is reconstructed with the HPTWTagger, while perfect reconstruction is assumed for the leptonically decaying $Z$ boson.
 In the right panel the $Z$ boson decays hadronically and is reconstructed with the HPTZTagger, while the leptonically decaying $W$ boson is reconstructed as described in the text.
 The relevant Standard Model backgrounds consist of $Z+\text{jets}$ (left) and $W+\text{jets}$ (right) respectively, and are depicted as black dashed lines.
 Event rates are given for $pp$ collisions at $\sqrt{s}=14\TeV$ and an integrated luminosity of $300\invfb$.}
 \label{fig:wztagger_resonance}
\end{figure}

The heavy $W'$ resonance is reconstructed as the vectorial sum of the tagged $W$ and $Z$ four-momenta.
Its invariant mass we plot in~\figref{wztagger_resonance} (l.h.s.), showing expected event rates for an integrated luminosity of 300\invfb.
For all benchmark masses, the signal peaks outnumber the SM background in the respective mass bins.

The second scenario is given by a leptonically decaying $W$ boson,
$pp\to W'^\pm\to W^\pm Z\to (l^\pm\nu)(jj)$
and we apply the HPTZTagger to the hadronic decay of the $Z$ boson.
The relevant Standard Model background is $p p\to W^\pm j$ in the highly-boosted regime, again generated in generator-level $\hat{p}_\perp$ bins in the range $[700\GeV, 2.5\TeV]$ and $\hat{p}_\perp\geq 2.5\TeV$.

To recover the four-momentum of the $W$ boson, we assume perfect reconstruction of the charged lepton and obtain the four-momentum of the invisible neutrino as follows.
Standard jets (C/A, $R=0.4$, $p_\perp\geq 30\GeV$) are used to determine the missing transverse momentum.
Jets in the vicinity of the charged lepton are discarded if $\Delta R_{j,l} < 0.5$.
The neutrino transverse momentum is then given by the negative of the summed jet transverse momentum.
Imposing zero invariant mass of the neutrino and fixing the $W$ mass $(p_l+p_\nu)^2=m_W^2$, we can determine the remaining components of the neutrino momentum.
Of the two solutions of the quadratic equation for $p_z$, we choose the one with smaller $\Delta R_{l\nu}$.
Due to imperfect knowledge of transverse momentum, there may be no real solution.
In this case we simply take the real part as $p_z$.

Candidate fat jets for $Z$ boson tagging are rejected if not well-separated from the decay products of the $W$ boson,
$\Delta R(j,\mathrm{lepton/neutrino}) < 0.6$.
We then obtain the tagged mass of the heavy $W'$ resonance from the reconstructed $W$ and $Z$ bosons, $m_{W'}^2 = (p_W + p_Z)^2$, see~\figref{wztagger_resonance} (r.h.s.).

As was done in \secref{res:top}, we evaluate the required integrated luminosity to exclude the three benchmark resonances using a cut and count analysis.
We select the two bins with largest significance ($S/\sqrt{B}$) for the reconstructed $W'$ with the HPTWTagger in \tabref{wtag}.
Using the HPTZTagger on the second scenario, we find less significant results, see \tabref{ztag}.

\begin{table}
\begin{center}
\begin{tabular}{c|c|c|c|c|c}
 Resonance & $m_{12}$ window (TeV) & $S/B$ & $S/\sqrt{B}$ & $\sigma$(fb) & $\sigma$ for $S/\sqrt{B}=2$  \\\hline
 $m_{W'}=3$ TeV & 2.9-3.1 & 10.1 & 11.9 & 0.139 & 0.0233 \\\hline
 $m_{W'}=4$ TeV & 3.9-4.1 & 15.8 & 5.1 & 0.0192 & 0.0075\\\hline
 $m_{W'}=5$ TeV & 4.9-5.1 & 17.3 & 1.6 & 0.00322 & 0.0040
\end{tabular}
\end{center}
\caption{Expected event rates for a heavy $W'$ reconstructed with the HPTWTagger at 300\invfb.}
\label{tab:wtag}
\end{table}

\begin{table}
\begin{center}
\begin{tabular}{c|c|c|c|c|c}
 Resonance & $m_{12}$ window (TeV) & $S/B$ & $S/\sqrt{B}$ & $\sigma$(fb) & $\sigma$ for $S/\sqrt{B}=2$  \\\hline
 $m_{W'}=3$ TeV & 2.9-3.1 & 3.80 & 10.6 & 0.455 & 0.086\\\hline
 $m_{W'}=4$ TeV & 3.9-4.1 & 2.78 & 2.92 & 0.0631 & 0.043\\\hline
 $m_{W'}=5$ TeV & 4.9-5.1 & 0.087 & 0.029 & 0.0108 & 0.743
\end{tabular}
\end{center}
\caption{Expected event rates for a heavy $W'$ reconstructed with the HPTZTagger at 300\invfb.}
\label{tab:ztag}
\end{table}

\begin{figure}[htb]
 \begin{center}
  \includegraphics[trim=5cm 14cm 3cm 4cm, clip=true, width=0.7\textwidth]{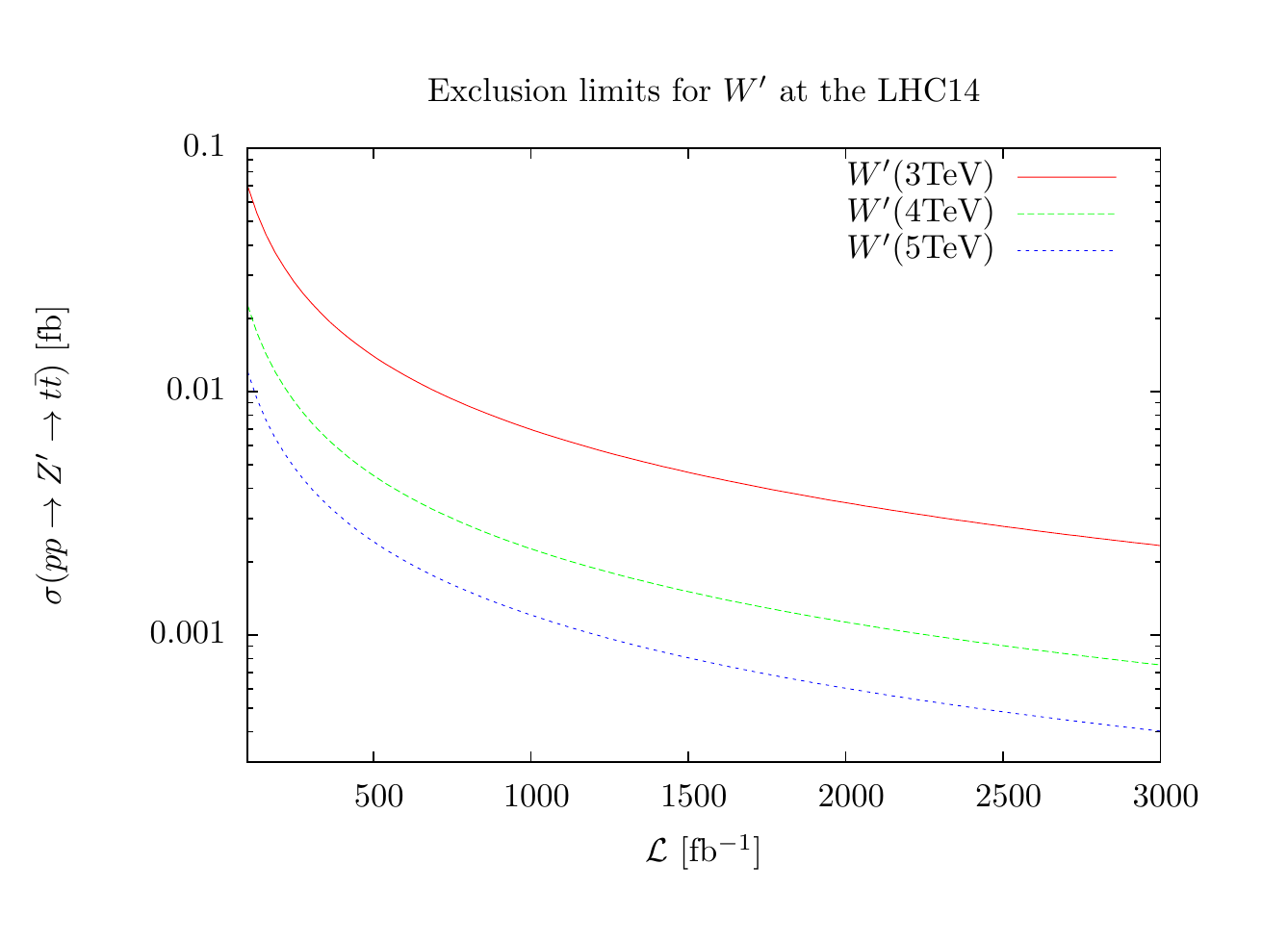}
 \end{center}
 \caption{Exclusion limits at 90\% CL  
 for the process $pp\to W'^\pm\to W^\pm Z$ at $\sqrt{s}=14\TeV$.
 The hadronically decaying $W$ boson is tagged and reconstructed with HPTWTagger.}
 \label{fig:wtagger_resonance_exclusion}
\end{figure}

Based on the results of the $pp\to W'^\pm\to W^\pm Z\to (jj)(l^+l^-)$ process, we give the required integrated luminosity to exclude a heavy $W'$ resonance at the LHC depending on the production cross-section in \figref{wtagger_resonance_exclusion}.

\section{Tagging highly boosted scalars at the LHC}
\label{sec:higgstag}

After the recent discovery of a Higgs boson, i.e.~the first electroweak-scale scalar resonance, the scalar sector has become the centre of the focus of both multi-purpose experiments and the theory community. Whether the Higgs boson is part of a minimal or non-minimal Higgs sector remains to be determined. While many popular models predict several scalar resonances, e.g.~2HDM or NHDM, their masses are a priori largely undetermined.

To give an example for the benefit of using a track-based tagger for scalar resonances at the LHC with $\sqrt{s}=14$ TeV, we will focus on the rare prompt decay of the Higgs boson into a $Z$ boson and a cp-odd light scalar $H \to Z A$. The mass of the cp-odd scalar is assumed to be $m_A = 2$ GeV. Hence, $A$ is likely to be highly boosted as $m_H/2 \gg m_A$. We assume $A$ to decay into gluons exclusively. To cope with large backgrounds we study the Higgs boson produced in association with a leptonically decaying $Z$ boson.

The signal we generate using \pythia and for the background process $ZZj$ we use \sherpa \cite{sherparefs}. Including decays of the $Z$ bosons to electrons or muons and imposing $p_{\perp,j}\geq 25$ GeV, we find a LO cross-section of $\sigma_{ZZj} =31.69$ fb.

\begin{figure}[htb]
 \begin{center}
  \includegraphics[width=0.7\textwidth]{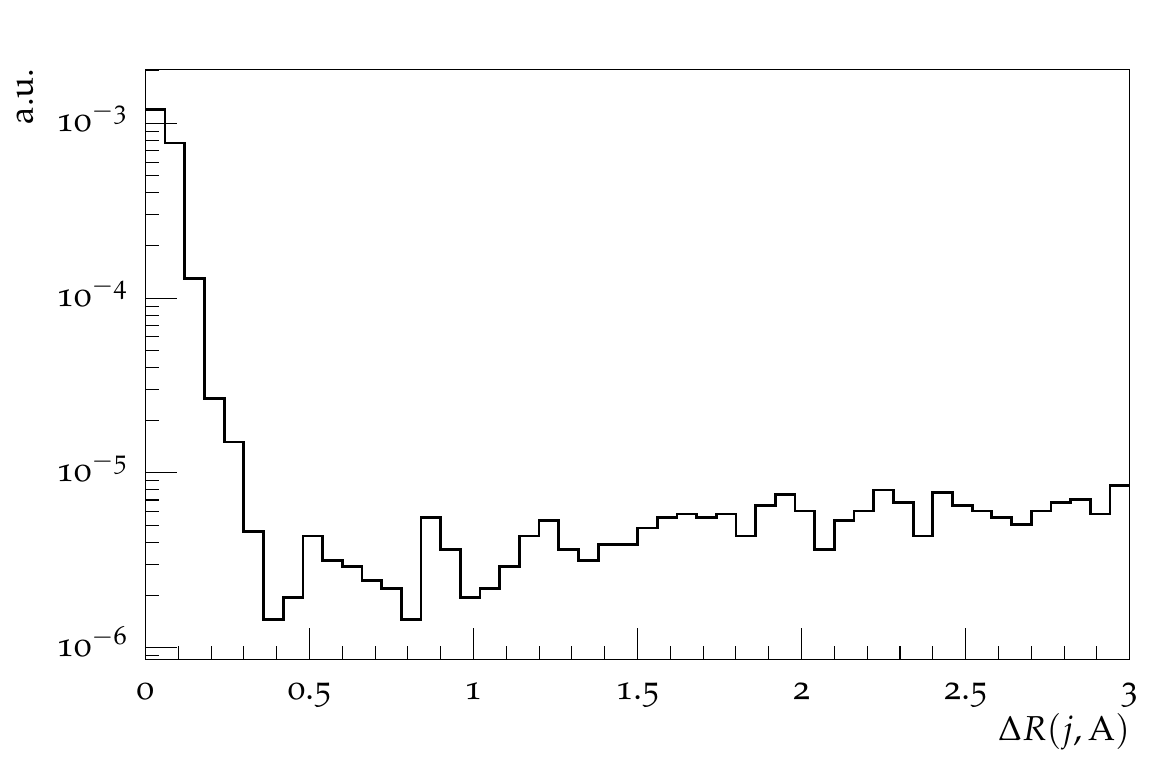}
 \end{center}
 \caption{Angular distance between the selected jet and the true cp-odd scalar $A$.}
 \label{fig:scalar:dr}
\end{figure}

We require exactly four leptons (electrons or muons) with $p_{\perp,l} \geq 10$ GeV. To pair the four leptons to two $Z$ bosons we minimize
\begin{equation}
\chi_{ZZ}^2 = \frac{(m_{l_i,l_j} - m_Z)^2}{\Delta_Z^2} + \frac{(m_{l_n,l_m} - m_Z)^2}{\Delta_Z^2}
\end{equation}
where $i \neq j \neq n \neq m$ and $\Delta_Z = 5$ GeV. We then require both lepton pairs to be in a mass window of $m_Z \pm 5$ GeV individually.
We then remove the leptons from the final state objects and use the remaining objects in $|\eta| < 5$  to cluster C/A jets with $R=0.4$ and $p_{\perp,j} \geq 30\GeV$. Next we need to identify the jet containing the decay products of the cp-odd scalar. Thus, we select this jet by minimizing
\begin{equation}
\chi_{jZ} = \min (\left | m_{j,Z_1} - m_H \right | , \left | m_{j,Z_2} - m_H \right | )
\end{equation}
and veto events where $\chi_{jZ,\mathrm{min}} > 15$ GeV.
We find that this requirement selects the jet with smallest distance to the cp-odd scalar efficiently, see \figref{scalar:dr}.
For the signal jet we expect a very narrow pencil-like substructure with a high localised energy density, similar to hadronic $\tau$ jets. To further remove background contamination we require at least two charged tracks associated with the jet and eventually use all charged tracks as constituents to construct $j_c$. We veto events if $m_{j_c} > 3$ GeV.

\begin{table}
\begin{center}
\begin{tabular}{c|c|c|c|c}
 & 2 $Z$ bosons & fat jet & $\chi_{jZ} < 15$~GeV & $n_{\mathrm{tracks}} > 1$ and $m_{j_c}<3$ GeV  \\\hline
 $(H \to ZA)Z$ & 0.513 & 0.296 & 0.221 & 0.131  \\\hline
 $ZZj$ & 0.497 & $1.61 \cdot 10^{-3}$ & $4.84 \cdot 10^{-4}$ & $5.43\cdot 10^{-5}$
\end{tabular}
\end{center}
\caption{Reconstruction efficiencies $\epsilon$ for analysis steps described in \secref{higgstag}.}
\label{tab:HZA}
\end{table}

Using the reconstruction efficiencies $\epsilon$ in \tabref{HZA}, with $\sigma_{HZ} =883.0~\mathrm{fb}$ and assuming BR$(A \to gg) = 1$ we can set a limit BR$(H \to ZA) < 0.001$ for 100 $\mathrm{fb}^{-1}$.

\section{Summary and outlook}
\label{sec:outlook}

The reconstruction of heavy resonances is of great importance for the success of the upcoming LHC runs and possible future colliders. When heavy resonances decay into electroweak-scale objects,
these objects are highly boosted.
The subsequent decay products are then very much collimated and inevitably merge to a single jet, unable to be separately resolved by the detector's calorimeter.
For the ATLAS and CMS detectors at the LHC, this angular resolution scale corresponds to heavy resonances with mass $\gtrsim 2\TeV$ when the decay products are around the electroweak scale.

We developed dedicated tagging algorithms for the hadronic three-prong decay of the top quark (HPTTopTagger) as well as for the hadronic two-prong decay of $W$ and $Z$ bosons (dubbed HPTWTagger and HPTZTagger).
Our algorithms show stable efficiencies and kinematic reconstruction up until boosts of several TeV.
The taggers combine the good energy resolution of the calorimeter with the very fine spatial resolution of the tracker.
As only charged particles leave a signal in the tracking detector, this partial information alone is insufficient.
We apply optimized substructure techniques on highly boosted fat jets and find good discrimination power against QCD-initiated jets, which form the dominant background.

Track-based tagging algorithms significantly increase the discovery reach of heavy resonances into the multi-TeV regime.
We showed that already in the early stage of run II at the LHC, a heavy $Z'$ gauge boson decaying into a pair of top quarks can be excluded up until $m_{Z'}=3\TeV$, resulting in a much stricter bound than current exclusion limits.
A future proton-proton collider with centre-of-mass energy $\sqrt{s}=100\TeV$ is capable of probing heavy resonances of multiples of $\mathcal{O}(10)$ TeV.
Applying the HPTWTagger and HPTZTagger to a search for a heavy charged boson $W'$, we find an exclusion reach up to $m_{W'}=4\TeV$ in the early stage of LHC run II.
Probing such large resonance masses is not possible with current tagging algorithms based solely on jets built with calorimeter information.

We also showed that the general arguments on spatial resolution do not apply exclusively to very heavy resonances by constraining the branching ratio of the 125 GeV Higgs boson into a $Z$ boson and a very light cp-odd scalar.
The essential quantity that determines whether or not charged tracks have to be taken into the picture is the mass ratio of the heavy resonance $X$ and the intermediate resonance $Y$ that ultimately decays into the observed jets.
Whenever $m_X/m_Y \sim 20$, reconstruction methods relying on calorimeter-based jets break down and track-based observables start to become indispensable.

\begin{acknowledgments}
We thank Sebastian Sch\"atzel for helpful discussions.
M.~Stoll was supported by the Program for Leading Graduate Schools, MEXT, Japan.
M.~Stoll is grateful for hospitality of the IPPP in Durham.

\end{acknowledgments}


\begin{thebibliography}{99}

\bibitem{Aad:2012tfa}
  G.~Aad {\it et al.}  [ATLAS Collaboration],
  Phys.\ Lett.\ B {\bf 716}, 1 (2012).

\bibitem{Chatrchyan:2012ufa}
  S.~Chatrchyan {\it et al.}  [CMS Collaboration],
  Phys.\ Lett.\ B {\bf 716}, 30 (2012).


\bibitem{Aad:2015pfa} 
  G.~Aad {\it et al.}  [ATLAS Collaboration],
  arXiv:1503.04430 [hep-ex].

\bibitem{Aad:2014xea} 
  G.~Aad {\it et al.}  [ATLAS Collaboration],
  Phys.\ Lett.\ B {\bf 743}, 235 (2015)
  [arXiv:1410.4103 [hep-ex]].


\bibitem{Khachatryan:2015sja} 
  V.~Khachatryan {\it et al.}  [CMS Collaboration],
  Phys.\ Rev.\ D {\bf 91}, no. 5, 052009 (2015)
  [arXiv:1501.04198 [hep-ex]].

\bibitem{Khachatryan:2015lwa} 
  V.~Khachatryan {\it et al.}  [CMS Collaboration],
  arXiv:1502.06031 [hep-ex].

\bibitem{Ellis:1993tq} 
  S.~D.~Ellis and D.~E.~Soper,
  Phys.\ Rev.\ D {\bf 48}, 3160 (1993)
  [hep-ph/9305266].

\bibitem{Catani:1993hr} 
  S.~Catani, Y.~L.~Dokshitzer, M.~H.~Seymour and B.~R.~Webber,
  Nucl.\ Phys.\ B {\bf 406}, 187 (1993).


\bibitem{ca_algo}
 Y.~L.~Dokshitzer, G.~D.~Leder, S.~Moretti and B.~R.~Webber,
  JHEP {\bf 9708}, 001 (1997);
 M.~Wobisch and T.~Wengler,
  arXiv:hep-ph/9907280.


\bibitem{Cacciari:2008gp} 
  M.~Cacciari, G.~P.~Salam and G.~Soyez,
  JHEP {\bf 0804}, 063 (2008)
  [arXiv:0802.1189 [hep-ph]].


\bibitem{LopezMateos:2011fta} 
  D.~Lopez Mateos,
  CERN-THESIS-2011-039.
  
  
\bibitem{thaler_wang}
 J.~Thaler, L.~-T.~Wang,
  JHEP {\bf 0807}, 092 (2008).

\bibitem{leandro1}
 L.~G.~Almeida, S.~J.~Lee, G.~Perez, I.~Sung and J.~Virzi,
  Phys.\ Rev.\  D {\bf 79}, 074012 (2009).


\bibitem{nsub}
 J.~Thaler, K.~Van Tilburg,
  JHEP {\bf 1103}, 015 (2011);
 J.~Thaler and K.~Van Tilburg,
  JHEP {\bf 1202}, 093 (2012).

\bibitem{treeless}
 M.~Jankowiak, A.~J.~Larkoski,
  JHEP {\bf 1106}, 057 (2011).

\bibitem{hopkins}
 D.~E.~Kaplan, K.~Rehermann, M.~D.~Schwartz and B.~Tweedie,
  Phys.\ Rev.\ Lett.\  {\bf 101}, 142001 (2008).

\bibitem{pruning1}
 S.~D.~Ellis, C.~K.~Vermilion and J.~R.~Walsh,
  Phys.\ Rev.\ D\ {\bf 80}, 051501  (2009).

\bibitem{cmstagger}
 CMS Collaboration,
  CMS-PAS-JME-09-001.


\bibitem{heptop}
  T.~Plehn, G.~P.~Salam and M.~Spannowsky,
  Phys.\ Rev.\ Lett.\  {\bf 104}, 111801 (2010);
  T.~Plehn, M.~Spannowsky, M.~Takeuchi, and D.~Zerwas,
  JHEP {\bf 1010}, 078 (2010);
  \url{http://www.thphys.uni-heidelberg.de/~plehn/}


\bibitem{Soper:2012pb}
  D.~E.~Soper and M.~Spannowsky,
  Phys.\ Rev.\ D {\bf 84}, 074002 (2011);
  D.~E.~Soper and M.~Spannowsky,
  Phys.\ Rev.\ D {\bf 87}, 054012 (2013).

  
  
\bibitem{hpttop}
  S.~Schaetzel and M.~Spannowsky,
  Phys.\ Rev.\ D {\bf 89} (2014) 014007
  [arXiv:1308.0540 [hep-ph]].




\bibitem{Katz:2010mr}
  A.~Katz, M.~Son and B.~Tweedie,
  JHEP {\bf 1103}, 011 (2011).



\bibitem{Larkoski:2015yqa} 
  A.~J.~Larkoski, F.~Maltoni and M.~Selvaggi,
  arXiv:1503.03347 [hep-ph].

\bibitem{Aad:2008zzm}
  G.~Aad {\it et al.}  [ATLAS Collaboration],
  JINST {\bf 3}, S08003 (2008).


\bibitem{Aad:2010ac}
  G.~Aad {\it et al.}  [ATLAS Collaboration],
  New J.\ Phys.\  {\bf 13}, 053033 (2011).



\bibitem{Aad:2012vm}
  G.~Aad {\it et al.}  [ATLAS Collaboration],
  Eur.\ Phys.\ J.\ C {\bf 73}, 2305 (2013).

\bibitem{Aad:2013gja}
  G.~Aad {\it et al.}  [ATLAS Collaboration],
  arXiv:1306.4945 [hep-ex].

\bibitem{Sjostrand:2007gs}
  T.~Sjostrand, S.~Mrenna and P.~Z.~Skands,
  Comput.\ Phys.\ Commun.\  {\bf 178}, 852 (2008)
  [arXiv:0710.3820 [hep-ph]].


\bibitem{fastjet}
 M.~Cacciari and G.~P.~Salam,
  Phys.\ Lett.\  B {\bf 641}, 57 (2006);
M.~Cacciari, G.~P.~Salam and G.~Soyez,
  Eur.\ Phys.\ J.\ C {\bf 72}, 1896 (2012);
 \url{http://fastjet.fr}


\bibitem{Ovyn:2009tx}
  S.~Ovyn, X.~Rouby and V.~Lemaitre,
  arXiv:0903.2225 [hep-ph].


\bibitem{Hocker:2007ht} 
  A.~Hocker, J.~Stelzer, F.~Tegenfeldt, H.~Voss, K.~Voss, A.~Christov, S.~Henrot-Versille and M.~Jachowski {\it et al.},
  PoS ACAT {\bf }, 040 (2007)
  [physics/0703039 [PHYSICS]].
\bibitem{chargedIR}
W.~J.~Waalewijn,
  Phys.\ Rev.\ D {\bf 86}, 094030 (2012);
H.~-M.~Chang, M.~Procura, J.~Thaler and W.~J.~Waalewijn,
  arXiv:1303.6637 [hep-ph].
H.~-M.~Chang, M.~Procura, J.~Thaler and W.~J.~Waalewijn,
  arXiv:1306.6630 [hep-ph];
  A.~J.~Larkoski and J.~Thaler,
  arXiv:1307.1699 [hep-ph].

\bibitem{Bahr:2008pv}
  M.~Bahr, S.~Gieseke, M.~A.~Gigg, D.~Grellscheid, K.~Hamilton, O.~Latunde-Dada, S.~Platzer and P.~Richardson {\it et al.},
  Eur.\ Phys.\ J.\ C {\bf 58}, 639 (2008).


\bibitem{ATLAS-CONF-2015-009}
ATLAS-CONF-2015-009, ATLAS collaboration

\bibitem{Wells:2008xg} 
  J.~D.~Wells,
  In *Kane, Gordon (ed.), Pierce, Aaron (ed.): Perspectives on LHC physics* 283-298
  [arXiv:0803.1243 [hep-ph]].

\bibitem{Dugan:1984hq} 
  M.~J.~Dugan, H.~Georgi and D.~B.~Kaplan,
  Nucl.\ Phys.\ B {\bf 254}, 299 (1985).

\bibitem{Schmaltz:2005ky} 
  M.~Schmaltz and D.~Tucker-Smith,
  Ann.\ Rev.\ Nucl.\ Part.\ Sci.\  {\bf 55}, 229 (2005)
  [hep-ph/0502182].

\bibitem{Bell:2010gi} 
  G.~Bell, J.~H.~Kuhn and J.~Rittinger,
  Eur.\ Phys.\ J.\ C {\bf 70}, 659 (2010)
  [arXiv:1004.4117 [hep-ph]].
  
\bibitem{Christiansen:2014kba} 
  J.~R.~Christiansen and T.~Sjöstrand,
  JHEP {\bf 1404}, 115 (2014)
  [arXiv:1401.5238 [hep-ph], arXiv:1401.5238].
  
\bibitem{Krauss:2014yaa} 
  F.~Krauss, P.~Petrov, M.~Schoenherr and M.~Spannowsky,
  Phys.\ Rev.\ D {\bf 89}, no. 11, 114006 (2014)
  [arXiv:1403.4788 [hep-ph]].
  
\bibitem{Krohn:2012fg} 
  D.~Krohn, M.~D.~Schwartz, T.~Lin and W.~J.~Waalewijn,
  Phys.\ Rev.\ Lett.\  {\bf 110}, no. 21, 212001 (2013)
  [arXiv:1209.2421 [hep-ph]].

\bibitem{bdrs}
  J.~M.~Butterworth, A.~R.~Davison, M.~Rubin and G.~P.~Salam,
  Phys.\ Rev.\ Lett.\  {\bf 100}, 242001 (2008).

\bibitem{sherparefs}
  T.~Gleisberg, S.~Hoeche, F.~Krauss, M.~Schonherr, S.~Schumann, F.~Siegert and J.~Winter,
  JHEP {\bf 0902} (2009) 007.
  T.~Gleisberg, S.~Hoeche
  JHEP {\bf 0812} (2008) 039






\end{thebibliography}
\end{document}